\documentclass{article}

\usepackage[square]{natbib}
\usepackage{amssymb}
\usepackage{amsthm}
\usepackage{fancyhdr}
\usepackage{authblk}
\usepackage[title]{appendix}
\usepackage{array}
\usepackage[tableposition=top]{caption}
\usepackage{dsfont}
\usepackage{graphicx}
\usepackage{mathtools}
\usepackage{makecell}
\usepackage{multicol}
\usepackage{multirow}
\usepackage{pgfplots}
\usepackage{tikz}
\usepackage{xcolor}
\usepackage{enumitem}
\setlist[itemize]{noitemsep}
\setlist[enumerate]{noitemsep}
\usepackage{arxiv}
\usepackage{booktabs}       
\usepackage{nicefrac} 

\usepackage{enumitem}
\usepackage{algorithm}
\usepackage{algpseudocode}
\usepackage{xpatch}

\usepackage[superscript,biblabel]{cite}

\usepackage[OT1]{fontenc}

\usepackage{url}
\usepackage{hyperref}

\usepackage[capitalise]{cleveref}

\usepackage{amsmath}
\usepackage{arxiv}

\pgfplotsset{
    jitter/.style={
        x filter/.code={\pgfmathparse{\pgfmathresult+(rnd-.5)*#1}}
    },
    jitter/.default=0.5
}

\usetikzlibrary{hobby, graphs, decorations.markings, arrows, shapes, positioning, mindmap, backgrounds, calc, matrix, patterns, bayesnet, patterns.meta}

\usepgfplotslibrary{statistics, fillbetween, external, groupplots}

\makeatletter
\xpatchcmd{\algorithmic}{\itemsep\z@}{\itemsep=0.8ex plus2pt}{}{}
\makeatother

\makeatletter
\renewcommand{\fnum@algorithm}{\fname@algorithm}
\makeatother
\algrenewcommand\algorithmicrequire{\textbf{Input:}}
\algrenewcommand\algorithmicensure{\textbf{Output:}}
\algnewcommand{\LeftComment}[1]{\Statex \(\triangleright\) #1}

\usepackage{pgfplots}  
\usepackage{pgfplotstable}  

\pgfplotsset{compat=1.4}

\usetikzlibrary{hobby, graphs, decorations.markings, arrows, shapes, positioning, mindmap, backgrounds, calc, matrix, patterns, bayesnet, patterns.meta}

\usepgfplotslibrary{statistics, fillbetween, external}

\newtheorem*{theorem*}{Theorem}

\newtheorem*{corollary*}{Corollary}

\newcommand{\nA}{N_{ \mathcal A }}

\newcommand{\nB}{N_{ \mathcal B }}

\newcommand{\GAik}{G^{ \mathcal A }_{i, k}}

\newcommand{\GBjk}{G^{ \mathcal B }_{j, k}}

\newcommand{\HAik}{H^{ \mathcal A }_{i, k}}

\newcommand{\HBjk}{H^{ \mathcal B }_{j, k}}

\usepackage{listings}
\usepackage{xcolor}

\definecolor{codebg}{RGB}{247,247,247}
\definecolor{codecomment}{RGB}{96,96,96}
\definecolor{codekw}{RGB}{0,60,130}
\lstdefinelanguage{R}{
  morekeywords={function,if,else,for,while,repeat,in,next,break,TRUE,FALSE,NULL,NA,library,return},
  sensitive=true,
  morecomment=[l]{\#},
  morestring=[d]{"},
  morestring=[d]{'}
}

\definecolor{firebrick}{HTML}{B22222}
\definecolor{coral1}{HTML}{FF7256}
\definecolor{goldenrod1}{HTML}{FFC125}
\definecolor{Rtwo}{HTML}{DF536B}
\definecolor{Rthree}{HTML}{61D04F}
\definecolor{Rfour}{HTML}{2297E6}
\definecolor{Rfive}{HTML}{28E2E5}
\definecolor{Rsix}{HTML}{CD0BBC}
\definecolor{Rseven}{HTML}{F5C710}
\definecolor{Reight}{HTML}{9E9E9E}

\begin{document}

\title{ FlexRL: an R package for Flexible Record Linkage and Linked Data Quality Assessment }

\renewcommand\Authsep{, }
\renewcommand\Authands{, and }

\author{Kayan\'{e} Robach\thanks{Corresponding Author: \texttt{k.c.robach@amsterdamumc.nl}}}

\author{Mark A.\ van de Wiel}

\author{Michel H.\ Hof}

\affil{Department of Epidemiology and Data Science, \protect\\ Amsterdam UMC location Vrije Universiteit Amsterdam, De Boelelaan 1117, 1081 HV Amsterdam, The Netherlands}
\affil{Amsterdam Public Health, Methodology, The Netherlands}

\maketitle

\abstract{\textit{FlexRL} is an \texttt{R} package for probabilistic record linkage and inference on linked data. It provides a method to link observations from the same entities across two data sources that lack a unique identifier but share partially identifying variables. This method fits a hierarchical latent variable model of the data generation process, rather than classifying pairs of records from the comparison of their values, and accommodates registration errors and variables changing over time. Since linkage may affect subsequent analyses, the package also provides tools to assess the linked data quality as functions of the linkage decision rule. Two false discovery proportion estimators are implemented: a probabilistic approach contingent on the record linkage model and an augmented approach relying on an auxiliary estimation using synthetic data. Several measures of divergence of the linked sample from the source population are provided. These tools are agnostic to the underlying record linkage model and can be applied to any method. Together, they connect the output of the linkage to the validity of the subsequent analyses. The implementation combines standard \texttt{R} functions and S3 methods with \texttt{Rcpp} computations, resulting in fast execution times and a low memory footprint.}

\keywords{Record Linkage, Inference}

\section{Introduction}\label{sec: intro package}

Researchers and statistical agencies increasingly combine data sources that do not share a unique identifier \citep{Cleaton_Splink_2024, Laidler_NHSE_2025, Kramer_Destatis_2026}. Record linkage connects records referring to the same entity across sources based on Partially Identifying Variables (PIVs, e.g. birth year, sex, postal code or product code, brand and category). Each PIV has limited discriminative power on its own and may be missing, incorrectly recorded or change over time. Record linkage underpins the production of official statistics, enables the reuse of administrative and registry data for research \citep{opensafely2020, perined_linkage_2021, cooneyhealthcare_2023} and improves data quality through linkage and deduplication.

Linked data sets become the study sample for downstream analyses, where linkage errors may propagate \citep{reflexRL_2020, Wang_Slawski_linked_2022, CoxLinkedData_2024, CRLrobachetal26}. Errors arise when links between entities that appear in both sources are missed or when records from distinct entities are falsely linked. These errors cannot be reduced simultaneously and may bias the conclusions drawn from linked data. A stricter decision rule reduces falsely linked pairs but misses more links. Practitioners therefore need not only record linkage methods, but also tools to quantify linkage errors and appraise their consequences for statistical analysis.

Two main probabilistic approaches to record linkage have developed through times. First, the traditional mixture model \citep{FS_RL_1969, jaro_RL_matching_1989, Larsen_mixture_1999, fortini_bay_RL_2001, sadinle_bay_bipartite_RL_2017} treats linkage as a classification problem between pairs of records that are similar and pairs that are not. Second, the hierarchical entity resolution framework \citep{tancredi_bay_RL_2011, steorts2015, smered2016, dblink_2021, marchantsteorts2023} models the data generation process, considering observed values as potentially distorted versions of latent true values. The latter framework uses all the information contained in the records and is therefore more flexible but more computationally demanding. It has been implemented in \texttt{R} for entity resolution (clustering records from a single file into entities) rather than for linkage between two data sources.

We address this gap with the \textit{FlexRL} \texttt{R} package, which makes two contributions. First, it implements the hierarchical record linkage model proposed by \citet{RLrobachetal25}. The model is fitted with a Stochastic Expectation Maximisation (StEM) algorithm that accounts for registration errors and temporal changes in PIVs. Its architecture and efficient \texttt{Rcpp} backend support scalability to large data sets and a wide range of applications: health monitoring studies, survey waves, census and administrative records or data from different retailers. Second, it provides tools to assess the quality of the linked data, which quantify the linkage error and sample selection. \textit{FlexRL} implements two estimators of the False Discovery Proportion (FDP): one derived from linkage scores \citep{belin_rubin_1995, measurementError_2019, fastLink_paper_2019} and one based on synthetic data \citep{FDProbachetal25}. \textit{FlexRL} measures the divergence of the linked sample from the original sources using standardised mean differences, overlap of supports, maximum mean discrepancy and agreement rates. These tools only need the linked pairs and, when available, their linkage scores. They therefore apply to the output of any linkage method and \textit{FlexRL} provides wrappers for several existing \texttt{R} packages.

The \textit{FlexRL} package is aimed at practitioners, statistical agencies and researchers who need to link data sources without a unique identifier or who want to measure the error and the sample selection induced by record linkage. It builds on three lines of work: the biases that linkage errors cause in inference \citep{Lahiri_Larsen_linked_2005, Tancredi_Liseo_linked_2015, reflexRL_2020, Wang_Slawski_linked_2022, CRLrobachetal26}; the need for linked data quality control in practice \citep{eval_linkage_17, datalinkagequal2020, linkage_bias_datasynthesis24, Laidler_NHSE_2025}; advances in record linkage methodology \citep{sadinle_bay_bipartite_RL_2017, marchantsteorts2023, RLrobachetal25}. By combining record linkage and diagnostics in a practical tool, it helps evaluating whether inference can be conducted on linked data.

Throughout this chapter, we motivate and illustrate the main components of \textit{FlexRL} on the 2012 and 2014 waves of the \href{https://www.bancaditalia.it/statistiche/tematiche/indagini-famiglie-imprese/bilanci-famiglie/distribuzione-microdati/index.html}{Survey on Household Income and Wealth} (SHIW), conducted and made available by the Bank of Italy for research purposes. As a secondary analysis of public data, no additional ethics approval or participant consent was required. \cref{sec: setup} introduces the notation, the modelling paradigms, the existing software and demonstrates the potential of \textit{FlexRL}. \cref{sec: model} presents the record linkage model implemented in \textit{FlexRL} and its estimation on a subset of the SHIW. On the resulting linked data, \cref{sec: inference} describes the diagnostic tools \textit{FlexRL} provides, and illustrates the impact of linkage errors on a socio-economic regression analysis. \cref{sec: scalability} demonstrates the scalability of the \textit{FlexRL} approach to record linkage on the SHIW. \cref{sec: discussion} concludes.

\section{The record linkage problem}\label{sec: setup}

\subsection{Setup and notation}\label{sec: notation}

We consider two overlapping samples \smash{$\mathcal A$} and \smash{$\mathcal B$} from the same population with \smash{$\nA$} and \smash{$\nB \geq \nA$} records. Both sources record the same \smash{$K$} categorical PIVs, indexed by \smash{$k = 1, \dots, K$} with \smash{$n_k$} possible values each. Index \smash{$i$} refers to a record in \smash{$\mathcal A$} and index \smash{$j$} to a record in \smash{$\mathcal B$}. For PIV \smash{$k$}, we denote by \smash{$\GAik$} the registered (observed) value and by \smash{$\HAik$} the latent true value in record \smash{$i \in \mathcal A$}. We define \smash{$\GBjk$} and \smash{$\HBjk$} analogously. Registered values differ from true values because of registration errors: missing values or mistakes (e.g. typographical errors). We write \smash{$\textbf{G}$} and \smash{$\textbf{H}$} for the collections of all registered and all true values. True values may also differ between the two data collections. This is the case for \textit{dynamic} PIVs like postal code, whereas \textit{stable} PIVs like birth year cannot change. For an entity recorded in both sources, the true values of stable PIVs are therefore identical in \smash{$\mathcal A$} and \smash{$\mathcal B$} and any disagreement between their registered values is due to registration errors. We elaborate on dynamic PIVs later.

The object of interest is the linkage matrix \smash{$\boldsymbol{\Delta} \in \{0,1\}^{\nA \times \nB}$}, with \smash{$\Delta_{i,j} = 1$} if records \smash{$i \in \mathcal A$} and \smash{$j \in \mathcal B$} refer to the same entity. We assume that there are no duplicates within a data source. Thus, each record forms a link with at most one record in the other data source: \smash{$\sum_j \Delta_{i,j} \leq 1$} for all \smash{$i$} and \smash{$\sum_i \Delta_{i,j} \leq 1$} for all \smash{$j$}. This is the one-to-one assignment constraint, also referred to as bipartite matching \citep{sadinle_bay_bipartite_RL_2017}. We denote by \smash{$\gamma \in (0,1)$} the probability that a record from \smash{$\mathcal A$} forms a link with a record from \smash{$\mathcal B$}, that is, the overlap between the sources as a fraction of the smaller data set.

Since record linkage is imperfect, the linked data contain two types of error: false positives (pairs declared linked while referring to distinct entities) and false negatives (true links that are missed). The non-links in the Cartesian product \smash{$\mathcal A \times \mathcal B$} are far more numerous than the links and are rarely of interest. The useful summaries are therefore the False Discovery Proportion (FDP) among the linked data (complement of precision or positive predictive value) and the sensitivity (also called recall, complement of miss rate). The two errors affect inference differently: falsely linked pairs contaminate the inference and missed links induce sample selection \citep{reflexRL_2020, CRLrobachetal26}.

\subsection{Modelling paradigms}\label{sec: paradigms}

Linkage methods can be grouped into four paradigms \citep{review_methods_2014, Binette_Steorts_2022}. Deterministic (rule based) linkage declares a link when records agree on a set of variables, exactly or up to a tolerance. Supervised methods train a classifier on pairs whose linkage status is known \citep{DeepMatcher_method_2018}; such labelled pairs are seldom available. The two remaining paradigms are unsupervised and probabilistic, and differ in what they model:

\subsubsection{Paradigm 1: classification of comparison vectors}

The traditional `Fellegi-Sunter' approach \citep{FS_RL_1969, bookoflife, Newcombe1959, Tepping1968} reduces the data to comparison vectors. For each pair \smash{$(i,j) \in \mathcal A \times \mathcal B$}, a vector \smash{$\boldsymbol{c}_{i,j} = (c_{i,j,1}, \dots, c_{i,j,K})$} records whether elements \smash{$\GAik$} and \smash{$\GBjk$} for \smash{$k = 1, \dots, K$} agree or how similar they are. The pairs of records are split into the set of matched \smash{$M$} and the set of unmatched \smash{$U$}. The comparison vectors follow a two components mixture:
\begin{equation}\label{eq: FS}
\begin{aligned}
P(\boldsymbol{c}_{i,j}) &= \gamma \; m(\boldsymbol{c}_{i,j}) + (1-\gamma) \; u(\boldsymbol{c}_{i,j}), \\
m(\boldsymbol{c}) &= P(\boldsymbol{c} \mid M), \\
u(\boldsymbol{c}) &= P(\boldsymbol{c} \mid U),
\end{aligned}
\end{equation}
where \smash{$\gamma$} in this paradigm is the proportion of links among all pairs. Under the assumption that the \smash{$K$} comparisons are independent given the linkage status, \smash{$m$} and \smash{$u$} factorise over \smash{$k \in \{1, \dots, K\}$}. The parameters are then estimated with an EM algorithm \citep{jaro_RL_matching_1989, Larsen_mixture_1999}. Pairs are classified by thresholding the ratios \smash{$m(\boldsymbol{c}_{i,j}) / u(\boldsymbol{c}_{i,j})$} or the posterior linkage probabilities. Bayesian versions place a prior on the linkage matrix \smash{$\boldsymbol{\Delta}$} to enforce the one-to-one assignment constraint \citep{fortini_bay_RL_2001, sadinle_bay_bipartite_RL_2017} or a prior on a partition of the records when there are more than two sources or duplicates \citep{multifilepartitioningrecordlinkage}. The computations are simple and fast, but memory intensive because they operate on \smash{$\nA \times \nB$} comparison vectors. Scaling to large data sets may therefore require blocking, which compares only pairs that agree on some variables \citep{blocking_methods_2014}. This paradigm treats pairs as independent although they share records, and the reduction to comparison vectors discards part of the information contained in the records.

\subsubsection{Paradigm 2: hierarchical modelling}

The second family models how the data were generated \citep{tancredi_bay_RL_2011, steorts2015, smered2016, dblink_2021, marchantsteorts2023, RLrobachetal25}. Registered values \smash{$\textbf{G}$} are seen as distorted versions of latent true values \smash{$\textbf{H}$}, which define the linkage structure \smash{$\boldsymbol{\Delta}$}. The joint model reads
\begin{equation}\label{eq: hier}
P(\textbf{G}, \textbf{H}, \boldsymbol{\Delta}) = P(\textbf{G} \mid \textbf{H}) \, P(\textbf{H} \mid \boldsymbol{\Delta}) \, P(\boldsymbol{\Delta}),
\end{equation}
with a model for registration errors \smash{$P(\textbf{G} \mid \textbf{H})$}, a model for underlying true values \smash{$P(\textbf{H} \mid \boldsymbol{\Delta})$} and a model for the linkage structure \smash{$P(\boldsymbol{\Delta})$}. Since records are never reduced to pairwise comparisons, the model can separate missing values, mistakes and changes over time, and it does not need blocking to fit in memory. However, it is more computationally demanding.

\subsection{Existing software}\label{sec: software}

In \texttt{Python}, \textit{recordlinkage} \citep{PYRL_de_bruin_j_2019}, \textit{Splink} \citep{splink_2022} and \textit{NameMatch} \citep{PYRL_namematch} implement Paradigm 1. We list the main \texttt{R} packages for the task considered here in \cref{tbl: packages}. Paradigm 2 is only available in \texttt{R} through methods for entity resolution \citep{smered16software, dblink_2021, marchantsteorts2023}. These methods cluster records from a single file, possibly containing duplicates, into entities.

\begin{table}[htbp]
\centering
\setlength{\tabcolsep}{3pt}
\begin{tabular}{@{}p{25mm} p{18mm} p{23mm} p{21mm} p{21.5mm} p{15mm} p{12.5mm}@{}}
\toprule
Package & Paradigm & Estimation & \makecell{Linkage\\structure} & \makecell{Record\\deduplication} & \makecell{PIV\\dynamics} & \makecell{Quality\\tools} \\
\midrule
\textit{fedmatch} 2.1.0 \citep{fedmatchmanual, fedmatch_article} & comparison vectors & push relabel maximum flow & many-to-many & no & no & no \\
\textit{diyar} 0.5.1 \citep{diyarmanual} & comparison vectors & multistage weights comparisons, fixed score level parameters & clusters & yes & no & no \\
\textit{reclin2} 0.6.0 \citep{reclin2manual} & comparison vectors, supervised & EM, machine learning & one-to-one post hoc & no & no & no \\
\textit{fastLink} 0.6.1 \citep{fastLink_paper_2019, fastLinkmanual} & comparison vectors & EM & one-to-one post hoc & no & no & no \\
\textit{BRL} 0.1.0 \citep{brlmanual, sadinle_bay_bipartite_RL_2017} & comparison vectors & Bayesian, bipartite prior, Gibbs sampler & one-to-one in model & no & no & no \\
\textit{multilink} 0.1.1 \citep{multilinkmanual, multifilepartitioningrecordlinkage} & comparison vectors & Bayesian, partition prior, Gibbs sampler & clusters, one-to-one in model & yes & no & no \\
\textit{FlexRL} 2.0.0 \citep{flexrlmanual} & hierarchical & stochastic EM & one-to-one in model & no & yes & yes \\
\bottomrule
\end{tabular}
\caption{Record linkage packages on \texttt{CRAN} compared with \textit{FlexRL}. The \textit{FlexRL} package diagnostic tools apply to all these packages.}
\label{tbl: packages}
\end{table}

To our knowledge, no \texttt{R} package implements Paradigm 2 for the linkage of two sources under the one-to-one assignment constraint or models the dynamics of PIVs over time. While the \textit{postlink} package \citep{postlink26} develops statistical tools for post-linkage data analysis, no record linkage package provides tools to assess the quality of the linked data and their suitability for inference.

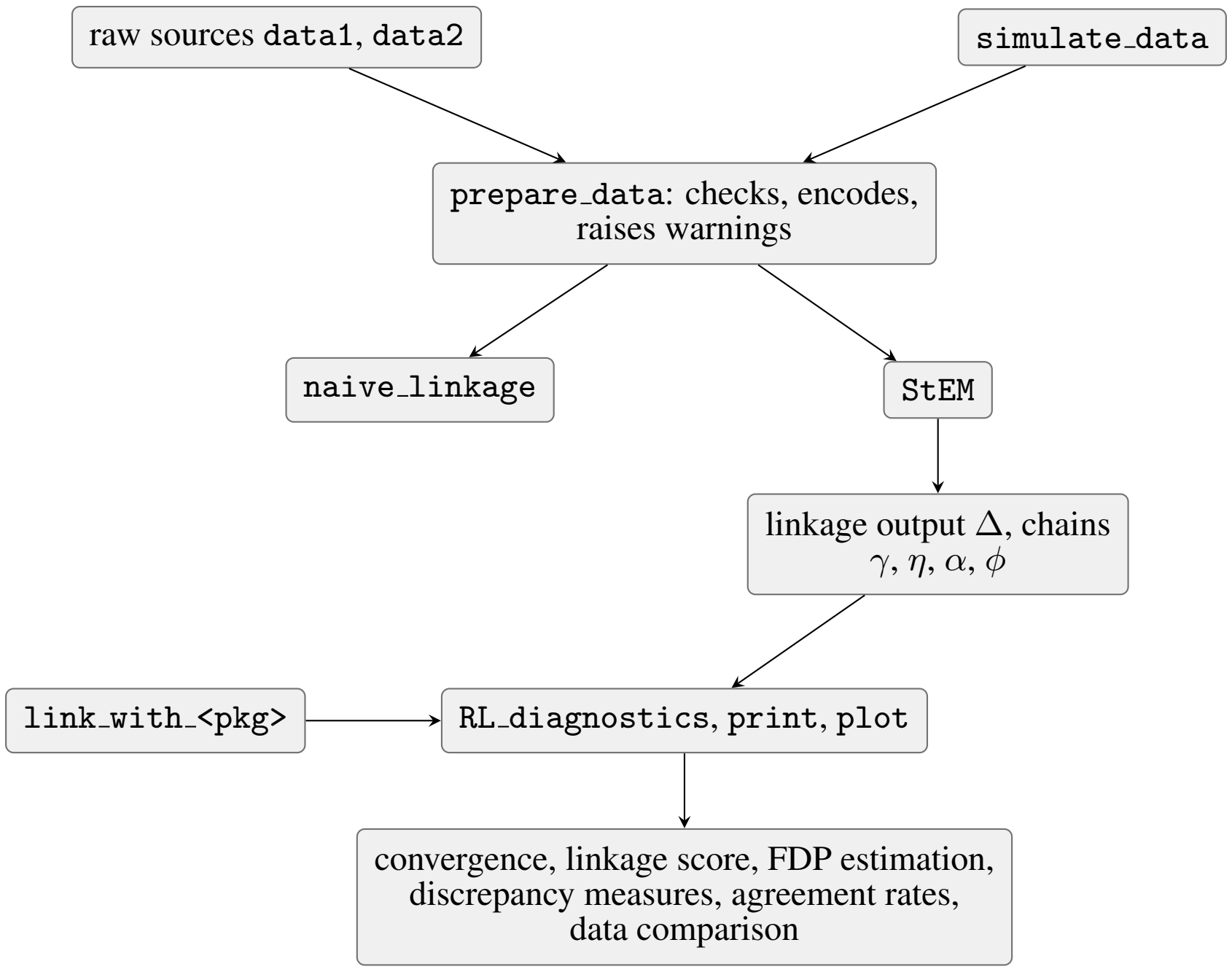
\begin{figure}[htbp]
\centering
\begin{tikzpicture}[>=stealth, font=\footnotesize,
  box/.style={draw=black!55, fill=black!6, rounded corners=2pt, align=center, inner sep=4.5pt}]
\node[box] (raw)   at (0,0)      {raw sources \texttt{data1}, \texttt{data2}};
\node[box] (sim)   at (7.4,0)    {\texttt{simulate\_data}};
\node[box] (prep)  at (3.7,-1.6) {\texttt{prepare\_data}: checks, encodes,\\ raises warnings};
\node[box] (naive) at (1.3,-3.2) {\texttt{naive\_linkage}};
\node[box] (fit)   at (6.0,-3.2) {\texttt{StEM}};
\node[box] (delta) at (6.0,-4.6) {linkage output $\Delta$, chains\\ $\gamma$, $\eta$, $\alpha$, $\phi$};
\node[box] (diag)  at (3.7,-6.2) {\texttt{RL\_diagnostics}, \texttt{print}, \texttt{plot}};
\node[box] (wrap)  at (-1.1,-6.2) {\texttt{link\_with\_<pkg>}};
\node[box] (alles) at (3.7,-7.8) {convergence, linkage score, FDP estimation,\\ discrepancy measures, agreement rates,\\ data comparison};
\draw[->] (raw) -- (prep);
\draw[->] (sim) -- (prep);
\draw[->] (prep) -- (naive);
\draw[->] (prep) -- (fit);
\draw[->] (fit) -- (delta);
\draw[->] (delta) -- (diag);
\draw[->] (wrap) -- (diag);
\draw[->] (diag) -- (alles);
\end{tikzpicture}
\caption{Workflow of \textit{FlexRL}.}
\label{fig: workflow}
\end{figure}

The \textit{FlexRL} package is available from the Comprehensive \texttt{R} Archive Network (\texttt{CRAN}) or from the package \texttt{GitHub} repository alongside documentation and vignettes describing its functionality. \textit{FlexRL} is also available as a module in \texttt{JASP 1.0} \citep{JASP2026}. The implementation of \textit{FlexRL} relies on \texttt{Rcpp} (potentially parallelised with \texttt{OpenMP}) for the backend samplers, \textit{Matrix} for the sparse linkage matrix, and other standard \texttt{R} packages. The diagnostic tools apply to any record linkage output in the required format, with built-in wrappers for \textit{BRL}, \textit{diyar}, \textit{fastLink}, \textit{fedmatch}, \textit{multilink} and \textit{reclin2}. \cref{fig: workflow} shows the workflow, which follows the structure of \cref{sec: model} and \cref{sec: inference}: the data are prepared, linked with \texttt{StEM()} and the linked data quality is assessed with \texttt{RL\_diagnostics()}. We elaborate on each component along this chapter.

\subsection{Demonstration}\label{sec: demonstration}

\subsubsection{Simulation study}

As a preliminary demonstration of the functionalities of \textit{FlexRL}, we simulate data with the package \texttt{simulate\_data()} function. This function synthesises two files with a given overlap, given rates of mistakes and missing values per PIV and file, and given hazard coefficients for structured PIVs, useful for exploring and comparing record linkage methods. Here, we generate two data sets of 250 and 300 records, with 200 entities common to both. Entities are represented by four independent discrete non-uniform PIVs. We distort the data in each file uniformly with 2\% of mistakes and 0.5\% of missing values; one of the PIVs undergoes temporal changes.

Note that this data generation process, on which all packages cited in \cref{tbl: packages} are evaluated, is in line with the hierarchical model \textit{FlexRL} fits and represents a simple setting. The fourth PIV follows a dynamic process whose hazard is piecewise constant over time and modulated by a Gaussian covariate. We fit these changes with an exponential survival model in \textit{FlexRL}, illustrating the benefits of modelling PIV dynamics, even under model misspecification, rather than treating them as registration errors. Later in this chapter, we evaluate all packages on real data for a fairer comparison.

\begin{table}[htbp]
\centering
\begingroup
\scriptsize
\setlength{\tabcolsep}{2.5pt}
\begin{tabular}{@{}l ccccccccc@{}}
\toprule
 & Naive & BRL & multilink & fastLink & reclin2 & diyar & fedmatch & \makecell{FlexRL\\(dynamics)} & \makecell{FlexRL\\(no dynamics)} \\
\midrule
TP & 118.40 (5.13) & 80.40 (5.04) & 80.60 (5.95) & 101.70 (6.48) & 114.10 (6.38) & 34.90 (20.24) & 122.10 (6.33) & 114.00 (7.92) & 87.50 (8.30) \\
FP & 83.10 (11.02) & 12.30 (3.23) & 14.20 (4.13) & 43.10 (5.59) & 68.30 (8.82) & 3.30 (2.91) & 125.80 (6.32) & 25.90 (5.15) & 19.00 (4.67) \\
FN & 81.60 (5.13) & 119.60 (5.04) & 119.40 (5.95) & 98.30 (6.48) & 85.90 (6.38) & 165.10 (20.24) & 77.90 (6.33) & 86.00 (7.92) & 112.50 (8.30) \\
sensitivity & 0.59 (0.03) & 0.40 (0.03) & 0.40 (0.03) & 0.51 (0.03) & 0.57 (0.03) & 0.17 (0.10) & 0.61 (0.03) & 0.57 (0.04) & 0.44 (0.04) \\
FDP & 0.41 (0.03) & 0.13 (0.03) & 0.15 (0.04) & 0.30 (0.03) & 0.37 (0.03) & 0.08 (0.05) & 0.51 (0.03) & 0.18 (0.03) & 0.18 (0.04) \\
time (secs) & 0.27 (0.04) & 0.97 (0.05) & 18.57 (1.48) & 1.29 (0.24) & 0.29 (0.01) & 1.17 (0.06) & 1.44 (0.01) & 4.59 (0.02) & 3.72 (0.01) \\
memory (Mb) & 1.04 (0.05) & 78.60 (10.08) & 105.68 (10.18) & 24.85 (4.42) & 41.52 (0.18) & 104.31 (10.84) & 42.14 (2.59) & 102.11 (9.03) & 70.18 (6.42) \\
\bottomrule
\end{tabular}
\endgroup
\caption{Linkage performance of competing record linkage methods on simulated data for the default linkage decision rule; see \cref{sec: scores}. We report average performance over 10 replications, with standard deviations in parentheses.}
\label{tbl: demonstration}
\end{table}

In \cref{tbl: demonstration}, we compare the methods presented in \cref{tbl: packages} to the \textit{FlexRL} record linkage algorithm with and without modelling PIV dynamics in terms of sensitivity and FDP. When appropriate, modelling PIV dynamics improves record linkage by increasing sensitivity (more true links are recovered) while keeping the FDP under control. The original methodological article provides a more comprehensive study of the behaviour of \textit{FlexRL} with and without modelling PIV dynamics \citep{RLrobachetal25}. \cref{tbl: demonstration} also shows the differences in computational time and in memory usage between the methods. More details on this demonstration showcasing the utility of accounting for PIV dynamics are given in the package vignette `Linkage and Diagnostics with FlexRL' available through \texttt{vignette(`FlexRL-RL-vignette', `FlexRL')} or the \href{https://robachowyk.github.io/FlexRL/index.html}{online documentation}.

The results presented in this chapter are obtained on a MacBook Pro (Apple M1 Max ARM chip architecture, 32GB memory, 8 cores) with \texttt{R} version 4.6.1. The script is available on the \href{https://github.com/robachowyk/FlexRL/tree/master/vignettes/article}{GitHub repository} of the package. The random seed is set to 1 in each code cell that produces output (some compared packages call \texttt{set.seed()} internally). The computations are robust to different operating systems (MacOS, Linux, Windows). We measure the memory usage using the \texttt{R} garbage collector. External allocations are therefore likely underestimated. All methods compared in this chapter rely on \texttt{Rcpp} or \texttt{C}, except for \textit{diyar} which is fully implemented in \texttt{R}. Time and memory usage may vary depending on the operating system, hardware, and system configuration.

\section{The hierarchical architecture}\label{sec: model}

\subsection{Data preparation and PIVs configuration}\label{sec: prepare}

The record linkage model requires each PIV \smash{$k$} to be categorical, encoded by integers \smash{$1, \dots, n_k$} with missing values coded as \smash{$0$}. The PIVs must cover a common support across the two sources. The smaller source should be labelled as \smash{$\mathcal A$}.

As a first step in the workflow represented in \cref{fig: workflow}, the \texttt{prepare\_data()} function which we will describe validates the PIVs configuration, labels the data sources and encodes the PIVs. When a unique identifier is available, like in simulation studies or in benchmark record linkage data such as the SHIW, it also returns the set of true links. This set may be used to evaluate linkage performance.

The function \texttt{naive\_linkage()} links records that agree exactly on all PIVs (a missing value matching any value when \texttt{na\_match = TRUE}), without enforcing the one-to-one constraint. When the true links are known, the performance of this deterministic linkage indicates the discriminative power of the PIVs and the difficulty of the task. The FDP of the `Naive' linkage reflects how many records share the same combination of PIV values.

The package distinguishes different kinds of PIVs, which determine how the components of the complete data likelihood of \cref{eq: flexrl} are used when fitting the model in \cref{sec: estimation}. Candidate pairs of records are formed from records whose latent true values agree on every PIV for which temporal changes are not modelled; whether a candidate pair is linked is then decided from its contribution to the complete likelihood. We elaborate on candidate pairs and linked pairs in \cref{sec: estimation}.

\begin{itemize}
\item \textit{Stable}: the latent true values do not change (e.g. birth year). Such PIV is used to form candidate pairs and its probability of mistake may be bounded (for example by \smash{$0.10$}) for algorithm stability \citep{RLrobachetal25}.
\item \textit{Dynamic}: the true values may change (e.g. postal code between two survey waves):
  \begin{itemize}
  \item \textit{Flexible}: changes cannot be modelled, typically because registration dates are not available. The PIV is still used to form candidate pairs, but its probability of mistake should not be bounded, so that the registration model absorbs both mistakes and changes.
  \item \textit{Structured}: changes are modelled by the survival function of \cref{eq: survival}. Such PIV is not used to form candidate pairs. Registration dates are needed in both sources and covariates for the hazard may be supplied. Mistakes cannot be distinguished from temporal changes, so the probability of mistake should be fixed (for example at \smash{$0.02$}), unless enough linked pairs have time gaps near zero, where changes are not plausible and mistakes can be estimated without confusion. Estimating the changes also requires variation in the time gaps among linked pairs.
  \end{itemize}
\end{itemize}

To illustrate the usage of the \textit{FlexRL} package throughout this chapter, we consider the Emilia-Romagna regional subset of the SHIW data. The record linkage task can be validated in this example where the linkage structure can be reconstructed. After preparation, the 2012 and 2014 waves cover 1,479 and 1,524 individuals, with 898 individuals observed in both waves. In \cref{sec: inference}, we analyse a regression of the expected age of retirement in 2014 on the expected age of retirement in 2012.

To link the two survey waves, we use the following variables: year and region of birth, sex, marital status, working status and branch of activity. The strong association between working status and branch of activity violates the PIV independence assumption of the method and may degrade the record linkage performance \citep{RLrobachetal25}. We thus combine these variables into a single variable `occupation'. Although marital status and occupation are `dynamic' PIVs, we do not have enough information to model their temporal changes and set them as `flexible' in the configuration:

\begin{lstlisting}[language=R]
PIVs_config <- list(BirthYear     = list(dynamics        = "stable",
                                         bound_mistakes  = c(0.10, 0.10),
                                         fix_mistakes    = c(NA, NA)),
                    Sex           = list(dynamics        = "stable",
                                         bound_mistakes  = c(0.10, 0.10),
                                         fix_mistakes    = c(NA, NA)),
                    MaritalStatus = list(dynamics        = "flexible",
                                         bound_mistakes  = c(NA, NA),
                                         fix_mistakes    = c(NA, NA)),
                    BirthRegion   = list(dynamics        = "stable",
                                         bound_mistakes  = c(0.10, 0.10),
                                         fix_mistakes    = c(NA, NA)),
                    Occupation    = list(dynamics        = "flexible",
                                         bound_mistakes  = c(NA, NA),
                                         fix_mistakes    = c(NA, NA)))

PIVs <- names(PIVs_config)
\end{lstlisting}

We prepare the two data sources for the \texttt{StEM()} record linkage algorithm with the \texttt{prepare\_data()} function, which takes as inputs the data sources and their labels \texttt{df2012}, \texttt{df2014}, `2012', `2014'; the PIVs configuration \texttt{PIVs\_config}; a logical \texttt{same\_mistakes} describing whether the parameters fitting the probability of mistakes among PIVs should be the same between data sources (\texttt{TRUE} when the data are sampled from a same source population); a potential record identifier variable \texttt{uniq\_id} when available; a logical \texttt{restrict\_support\_intersection} to harmonise the support of the PIVs in the observed data. In the application, records with birth years outside of the range shared by the sources are excluded.

\begin{lstlisting}[language=R]
prep_data <- prepare_data(df2012, df2014, "2012", "2014", PIVs_config,
                          same_mistakes = TRUE, uniq_id = "ID")
\end{lstlisting}

{\small
\begin{verbatim}
`2014' is the larger source, saved as B; `2012' saved as A.
\end{verbatim}
}

This function outputs a list suited for \texttt{StEM()}, where data sources have been encoded and labelled as \texttt{encodedA} and \texttt{encodedB}. It additionally includes the number of unique values per PIV \texttt{n\_values}, the validated \texttt{PIVs\_config}, the logical \texttt{same\_mistakes} and a data set of indices of true links \texttt{true\_pairs} when \texttt{uniq\_id} was provided. It alerts the user to potential issues, including inconsistent configurations, strongly associated PIVs and values outside the common support. For instance, the following warnings appear on the Emilia-Romagna subsets of the 2012 and 2014 waves of the SHIW data:

{\small
\begin{verbatim}
PIV `BirthYear': values out of the common support in data1 [1920, 1912]
in data2 [2013, 2014].
\end{verbatim}
}

{\small
\begin{verbatim}
PIVs `Occupation' and `Sex' are associated (Cramer's V = 0.32 in data1,
0.35 in data2); FlexRL assumes conditional independence between PIVs,
consider merging or dropping one.
\end{verbatim}
}

Among the true links, known in this application, there are some registration errors or temporal changes in the PIVs. As a consequence the true links do not necessarily carry the exact same information in their records. With the \texttt{RL\_agreement} function, the user can see the proportion of agreements among pairs of records (for example among linked pairs of records or, in the case presented below, among true links).

{\small
\begin{verbatim}
BirthYear   Sex   MaritalStatus   BirthRegion   Occupation
     0.99  1.00            0.98          0.99         0.82
\end{verbatim}
}

In \cref{tbl: illustration}, we compare the record linkage methods mentioned in \cref{tbl: packages} on this regional subset of the SHIW data. Memory usage is of the same order of magnitude on these relatively small data sets for all methods. The figures and tables based on the SHIW data reflect a single random realisation. They depend on the seed we use in the script and on the seed some other packages use internally.

\begin{table}[htbp]
\centering\small
\setlength{\tabcolsep}{3.2pt}
\begin{tabular}{@{}l cccccccc@{}}
\toprule
 & Naive & BRL & multilink & fastLink & reclin2 & diyar & fedmatch & FlexRL \\
\midrule
TP & 710.00 & 158.00 & 160.00 & 467.00 & 721.00 & 20.00 & 510.00 & 297.00 \\
FP & 3,494.00 & 24.00 & 72.00 & 850.00 & 339,232.00 & 0.00 & 969.00 & 103.00 \\
FN & 188.00 & 740.00 & 738.00 & 431.00 & 177.00 & 878.00 & 388.00 & 601.00 \\
sensitivity & 0.79 & 0.18 & 0.18 & 0.52 & 0.80 & 0.02 & 0.57 & 0.33 \\
FDP & 0.83 & 0.13 & 0.31 & 0.65 & 1.00 & 0.00 & 0.66 & 0.26 \\
time (secs) & 0.37 & 17.78 & 1,118.04 & 33.69 & 0.84 & 11.04 & 8.88 & 22.57 \\
memory (Mb) & 8.30 & 523.20 & 1,901.90 & 2,436.20 & 1,180.20 & 699.50 & 213.20 & 263.20 \\
\bottomrule
\end{tabular}
\caption{Comparison of record linkage methods on the SHIW in Emilia-Romagna (for the default linkage decision rule, see \cref{sec: scores}).}
\label{tbl: illustration}
\end{table}

\subsection{StEM estimation}\label{sec: estimation}

The \textit{FlexRL} hierarchical model for record linkage adapts \cref{eq: hier} to two sources under the one-to-one assignment constraint. For non-linked records, the true values are independent draws from the population distribution. For a linked pair \smash{$(i,j)$}, \smash{$\HBjk$} is a copy of \smash{$\HAik$} except when \smash{$k$} is a dynamic PIV, whose values may have changed between the two registrations. The registered values are versions of the true values distorted by registration errors.

With parameters \smash{$\boldsymbol{\theta} = (\gamma, \boldsymbol{\eta}, \boldsymbol{\alpha}, \boldsymbol{\phi})$}, the complete data likelihood factorises as
\begin{equation}\label{eq: flexrl}
\begin{aligned}
\mathcal{L}_{\boldsymbol{\theta}} \big( \textbf{G}^{\mathcal{A}}, \textbf{G}^{\mathcal{B}}, \textbf{H}^{\mathcal{A}}, \textbf{H}^{\mathcal{B}}, \boldsymbol{\Delta} \big)
= {} & \underbrace{\mathcal{L}_{\boldsymbol{\phi}^{\mathcal A}} \big( \textbf{G}^{\mathcal{A}} \mid \textbf{H}^{\mathcal{A}} \big)
\mathcal{L}_{\boldsymbol{\phi}^{\mathcal B}} \big( \textbf{G}^{\mathcal{B}} \mid \textbf{H}^{\mathcal{B}} \big)}_{\text{registration errors}} \\
& \underbrace{\mathcal{L}_{\boldsymbol{\alpha}, \boldsymbol{\eta}} \big( \textbf{H}^{\mathcal{B}} \mid \textbf{H}^{\mathcal{A}}, \boldsymbol{\Delta} \big)}_{\text{changes over time}} \;
\underbrace{\mathcal{L}_{\boldsymbol{\eta}} \big( \textbf{H}^{\mathcal{A}} \big)}_{\text{true values}} \\
& \underbrace{\mathcal{L}_{\gamma} \big( \boldsymbol{\Delta} \big)}_{\text{linkage}} .
\end{aligned}
\end{equation}
The parameter \smash{$\boldsymbol{\phi}$} governs the possible distortion mechanisms (mistakes, missing values) at the source of the differences between registered and latent true values. The parameter \smash{$\boldsymbol{\alpha}$} shapes the model for temporal changes between latent true values in data sources depending on the chosen `structure' (see the different survival models for \cref{eq: survival}). The parameter \smash{$\boldsymbol{\eta}$} characterises the distributions of the PIVs. The parameter \smash{$\gamma$} represents the probability that an observation in \smash{$\mathcal A$} forms a link with an observation in \smash{$\mathcal B$}. Each component of \cref{eq: flexrl} and each parameter is developed in the original methodological paper \citep{RLrobachetal25}.

Notably, the probability that true values have not changed between data collections is modelled with a survival function. By default, the package uses an exponential model with proportional hazards. For a dynamic \textit{structured} PIV \smash{$k$} whose changes are modelled and a linked pair \smash{$(i,j)$} with time gap \smash{$t_{i,j} = |t^{\mathcal B}_j - t^{\mathcal A}_i|$}:
\begin{equation}\label{eq: survival}
P\big(\HBjk = \HAik \mid t_{i,j}, \boldsymbol{x}_{i,j}, \Delta_{i,j} = 1\big) = S_{\boldsymbol{\alpha}_k}(\boldsymbol{x}_{i,j}, t_{i,j}) = \exp\big\{ - \exp(\boldsymbol{x}_{i,j}^{T} \boldsymbol{\alpha}_k) \, t_{i,j} \big\},
\end{equation}
where \smash{$\boldsymbol{x}_{i,j}$} contains an intercept and, optionally, covariates from either data source that explain the hazard of change (for example age or income for a relocation) and \smash{$\boldsymbol{\alpha}_k$} contains the log hazard coefficients. The user can choose another survival model, such as a Weibull, a Gompertz, a piecewise constant hazard or any custom model.

The likelihood of the observed data is obtained by summing \cref{eq: flexrl} over all values of the latent \smash{$\textbf{H}$} and \smash{$\boldsymbol{\Delta}$}. \textit{FlexRL} maximises it with the \texttt{StEM()} function, a Stochastic Expectation Maximisation (StEM) algorithm \citep{celeux_diebolt_SEM}. In a classical EM, the E-step computes the conditional expectation of the complete likelihood given the observed data. In the stochastic version, this expectation is approximated by Monte Carlo: a Gibbs sampler simulates the latent variables given the current parameters. The averages of the summary statistics collected after burn-in estimate the conditional expectation from which the M-step derives the parameter updates. \cref{alg: stem} summarises the procedure, detailed hereafter.

\begin{algorithm}
\caption{Stochastic EM for record linkage.}\label{alg: stem}
\begin{algorithmic}
\Require encoded sources \smash{$\textbf{G}^{\mathcal A}$, $\textbf{G}^{\mathcal B}$}; PIV configuration; \smash{$N_{\text{StEM}}$} iterations (burn-in \smash{$N0_{\text{StEM}}$}); \smash{$N_{\text{Gibbs}}$} iterations (burn-in \smash{$N0_{\text{Gibbs}}$})
\State Initialise \smash{$\boldsymbol{\theta} = (\gamma, \boldsymbol{\eta}, \boldsymbol{\alpha}, \boldsymbol{\phi})$}
\For{StEM iter \smash{$= 1, \dots, N_{\text{StEM}}$}}
    \State \smash{$\boldsymbol{\Delta} \gets 0$}
    \For{Gibbs iter \smash{$= 1, \dots, N_{\text{Gibbs}}$}} \Comment{stochastic E-step}
        \State draw \smash{$\textbf{H}^{\mathcal A}, \textbf{H}^{\mathcal B} \mid \textbf{G}^{\mathcal A}, \textbf{G}^{\mathcal B}, \boldsymbol{\Delta}, \boldsymbol{\theta}, \textbf{x}, \textbf{t}$}
        \State form candidate pairs \smash{$(i,j)$} such that \smash{$\HAik = \HBjk$} for all stable or flexible \smash{$k$}
        \State for each non-linked candidate pair: propose to link
        \State for each linked candidate pair: propose to break the link
        \State accept each proposal with the likelihood of the model
        \If{Gibbs iter \smash{$> N0_{\text{Gibbs}}$}}
            \State accumulate summary statistics:
            \State number of linked pairs,
            \State latent true values distribution,
            \State temporal changes in linked \smash{$(\HAik = \HBjk, \textbf{x}, \textbf{t})$},
            \State registration errors in linked and non-linked \smash{$(\textbf{G} = \textbf{H})$}
        \EndIf
    \EndFor
    \State average the accumulated summary statistics \Comment{estimate \smash{$\mathbb E[ \textbf{H}, \boldsymbol{\Delta} \mid \textbf{G}; \boldsymbol{\theta} ]$}}
    \State update and store \smash{$\boldsymbol{\theta} = (\gamma, \boldsymbol{\eta}, \boldsymbol{\alpha}, \boldsymbol{\phi})$} \Comment{M-step}
\EndFor
\State \smash{$\widehat{\boldsymbol{\theta}} \gets$} average of \smash{$\boldsymbol{\theta}$} over StEM iter \smash{$> N0_{\text{StEM}}$}
\State run further Gibbs iterations at \smash{$\widehat{\boldsymbol{\theta}}$} to build \smash{$\widehat{\boldsymbol{D}}$}:
\State \smash{$\widehat{D}_{i,j} \gets$} proportion of iterations in which \smash{$(i,j)$} is linked
\end{algorithmic}
\end{algorithm}

\subsubsection{Initialisation}

The probability \smash{$\gamma$} is drawn uniformly in \smash{$(0.2, 0.8)$}, the distribution of each PIV \smash{$k$}, \smash{$\boldsymbol\eta_k$}, are uniform over the \smash{$n_k$} values, the probabilities of mistakes are drawn in \smash{$(0.03, 0.20)$} and the probabilities of missing values are set to the observed proportions. For structured PIVs, the log baseline hazard is drawn between \smash{$0.01$} and \smash{$1$} per time unit and the covariate coefficients near \smash{$0$}. The user can supply other starting values.

\subsubsection{Stochastic E-step}

Given \smash{$\boldsymbol{\theta}$}, the Gibbs sampler alternates two blocks. First, the latent true values are drawn. For a non-linked record \smash{$\HAik$} or \smash{$\HBjk$} is drawn from \smash{$\boldsymbol\eta_k$} weighted by the registration model. For a linked pair \smash{$(\HAik, \HBjk)$} is drawn jointly among the \smash{$n_k \times n_k$} possible values, from \smash{$\boldsymbol\eta_k$} weighted by the registration model and the survival model.

Second, the linkage is updated. Candidate pairs are formed from matching sampled latent true values: records \smash{$i \in \mathcal A$} and \smash{$j \in \mathcal B$} are candidates when \smash{$\HAik = \HBjk$} for every non-structured PIV \smash{$k$} (stable or flexible). For each candidate pair \smash{$(i,j)$}, we evaluate the log likelihood contribution to \cref{eq: flexrl} when the records are linked, \smash{$\log \mathcal L^{\text{linked}}_{i,j}$}, and when they are not, \smash{$\log \mathcal L^{\mathcal A}_i + \log \mathcal L^{\mathcal B}_j$}. A sequential sweep over the candidate pairs then proposes to create or break links. A proposal is accepted by comparing the proposed `new' state of \cref{eq: flexrl} to the current `old' state with probability \smash{$\mathcal L_{\text{new}} / (\mathcal L_{\text{old}} + \mathcal L_{\text{new}})$}; the two states differ by \smash{$\log \mathcal L^{\text{linked}}_{i,j} - \log \mathcal L^{\mathcal A}_i - \log \mathcal L^{\mathcal B}_j$} and by the contributions from changes in the linkage matrix.

Over the Gibbs iterations, the sampler stochastically explores the latent space of possible linkage matrices and PIVs true values. At each iteration after burn-in we store the number of pairs that were linked, the distribution of latent true values, the temporal changes (disagreements between the linked records latent true values) and the registration errors (disagreements between \smash{$\textbf{G}$} and \smash{$\textbf{H}$})---summary statistics required for the M-step.

Restricting the sweep to candidate pairs keeps the algorithm light in memory: likelihood contributions are computed only for these pairs and for the remaining non-linked records in each data source, which together represent the estimated unique entities in the population.

\subsubsection{M-step}

The summary statistics averaged over the Gibbs draws after burn-in give the parameter updates. Thanks to the factorisation of \cref{eq: flexrl}, \smash{$\gamma$}, \smash{$\boldsymbol{\eta}$} and \smash{$\boldsymbol{\phi}$} have separate closed form updates \citep{RLrobachetal25}, and \smash{$\boldsymbol{\alpha}$} is obtained by numerical optimisation. Optional bounds on the probabilities of a mistake are applied at this stage. A warning is displayed when a bound is reached in many iterations, suggesting that the bound is too tight or that the PIV is dynamic.

\subsection{Linkage scores and decision rule}\label{sec: scores}

Based on the StEM chains of estimates, we build an estimate of the set of parameters, \smash{$\widehat{\boldsymbol{\theta}}$}, by averaging the post burn-in elements in the sequence. These are used to sample further linkage matrices that are averaged to return a probabilistic estimate of \smash{$\boldsymbol{\Delta}$} as a final output, namely \smash{$\widehat{\boldsymbol{D}}$}, whose entries \smash{$\widehat{D}_{i,j}$} are the posterior probabilities for the pair \smash{$(i,j)$} to be linked, which we call the linkage scores. A decision rule on each entry \smash{$\xi \in [0.5,1)$} then defines the final set of linked data \smash{$\mathcal{D} = \{(i,j) \,;\, \widehat{D}_{i,j} > \xi\}$}. Since the scores of a record sum to at most one, a threshold of at least \smash{$0.5$} ensures the one-to-one assignment \citep{tancredi_bay_RL_2011, sadinle_bay_bipartite_RL_2017}. The choice of \smash{$\xi$} is the subject of the upcoming \cref{sec: inference}.

Note that probabilistic record linkage methods estimate \smash{$\boldsymbol{\Delta}$} in comparable ways, by classifying pairs according to their linkage scores. Although these scores may be on different scales (e.g. likelihood ratios, log odds, posterior probabilities), they can be mapped to a probability scale. Hereafter, we take linkage scores to be posterior linkage probabilities in \smash{$[0,1]$} and treat all the methods listed in \cref{tbl: packages} in the same way. Accordingly, the default decision rule is \smash{$\xi=0.5$}. The `FDP versus linkage decision rule' curves (see \cref{fig: tradeoff}) provide a more comprehensive assessment of record linkage performance.

\subsection{Record linkage with \texorpdfstring{\texttt{StEM()}}{StEM()}}\label{sec: stem}

{\sloppy
\texttt{StEM()} runs \cref{alg: stem}. Its main arguments are:

\begin{itemize}
\item \texttt{data}: a list, typically the output of \texttt{prepare\_data()}, with the encoded sources \texttt{encodedA} and \texttt{encodedB}, the number of distinct values of each PIV \texttt{n\_values} (excluding missing values), a logical indicating whether both sources share the registration error parameters \texttt{same\_mistakes}, and the PIVs configuration \texttt{PIVs\_config}.
\item \texttt{StEM\_iter}, \texttt{StEM\_burnin}, \texttt{gibbs\_iter}, \texttt{gibbs\_burnin}: the numbers of iterations of the outer StEM algorithm and of the inner Gibbs sampler, and their burn-in periods. \texttt{gibbs\_burnin = 0} sets the burn-in automatically: the sampler runs until the number of links changes by fewer than 10 for 10 consecutive iterations, with a maximum of 500 iterations.
\item \texttt{threads}: the number of OpenMP threads used by the \texttt{Rcpp} samplers, default 4. The argument is ignored on systems where the package is built without \texttt{OpenMP}.
\item \texttt{model\_dynamics}: the survival model for structured PIVs, by default the exponential model of \cref{eq: survival}. It is only used if a PIV is declared structured in \texttt{PIVs\_config}.
\item \texttt{n\_post\_sample}: the number of posterior draws used to compute the linkage scores.
\item Starting values for \smash{$\gamma$} and for the agreement probabilities \smash{$1-\boldsymbol\phi$} can be supplied.
\item For long runs, \texttt{new\_directory} and \texttt{save\_info\_iter = TRUE} save the environment after each StEM iteration, and \texttt{music\_on = TRUE} plays a short tune when the algorithm finishes running.
\end{itemize}

The function returns \texttt{Delta}, a data frame with columns \texttt{i} (indices in \smash{$\mathcal A$}), \texttt{j} (indices in \smash{$\mathcal B$}) and \texttt{x} (scores \smash{$\widehat{D}_{i,j}$}) for every pair with a positive score, and the four chains: \texttt{gamma} (a vector of length \texttt{StEM\_iter}), \texttt{eta} (one \texttt{StEM\_iter} \smash{$\times$} \smash{$n_k$} matrix per PIV), \texttt{alpha} (one matrix per structured PIV, with \texttt{StEM\_iter} rows and one column per parameter of the survival model) and \texttt{phi} (one \texttt{StEM\_iter} \smash{$\times$} 4 matrix per PIV: agreement in \smash{$\mathcal A$}, in \smash{$\mathcal B$}, missing in \smash{$\mathcal A$}, in \smash{$\mathcal B$}).

We run the \textit{FlexRL} \texttt{StEM} algorithm for record linkage as follows:

\begin{lstlisting}[language=R]
fit <- StEM(data = prep_data)
\end{lstlisting}

\subsection{Computational aspects}\label{sec: computation}

The two latent variable samplers are implemented with \texttt{Rcpp}. To draw the true values, \texttt{sampleH()} goes through the \smash{$n_k^2$} joint values once per PIV \smash{$k$} for linked pairs and through the \smash{$n_k$} values once per PIV \smash{$k$} for non-linked records. \texttt{sampleD()} performs the sequential sweep over candidate pairs, recording the current linkage in two vectors that give, for each record of \smash{$\mathcal A$} and of \smash{$\mathcal B$}, the index of the record it is linked to (or a negative value when not linked). Reading and updating the linkage takes a fixed number of operations, so the sweep is linear in the number of candidate pairs and the memory is linear in the number of records \smash{$\nA + \nB$} rather than in \smash{$\nA \times \nB$}.

Candidate pairs are found by hashing the pattern of sampled true values on the non-structured PIVs, an operation linear in \smash{$\nA + \nB$}. Their contributions to the complete data likelihood are computed in a single pass over the records and the candidate pairs, implemented in \texttt{Rcpp}. Built-in survival models are evaluated in \texttt{Rcpp}; custom survival functions keep their \texttt{R} implementation.

The single pass over the candidate pairs likelihood contributions can be parallelised over \texttt{OpenMP} threads, each processing a disjoint set of candidates. Without \texttt{OpenMP} the code runs sequentially. All random draws occur upstream, so the threads never call the \texttt{R} random number generator. Results are therefore identical regardless of the number of threads, and \texttt{set.seed()} ensures reproducibility.

Running time grows with the number of records; the number of candidate pairs, larger when PIVs have few levels; the number of PIVs and their numbers of levels; the number of structured PIVs, whose survival model is estimated at each iteration; the number of iterations; and decreases with the number of threads.

\section{Inference on linked data}\label{sec: inference}

\subsection{Impact of linkage errors on linear regression}\label{sec: linkage error}

Suppose we are interested in estimating the association between \smash{$X$} recorded in \smash{$\mathcal A$} and \smash{$Y$} recorded in \smash{$\mathcal B$}, for instance through the slope \smash{$\beta$} of the linear regression \smash{$Y = X\beta + \varepsilon$}, with \smash{$\mathbb E[\varepsilon]=0$} and \smash{$\mathbb E[X \varepsilon]=0$}, so that
\begin{equation}\label{eq: beta-hat-0}
\widehat{\beta} = \operatorname{Cov}(X,Y) / \operatorname{Var}(X).
\end{equation}
The target is \smash{$\beta$} in the population of entities present in both sources. The analyst only observes the linked data \smash{$\mathcal{D} = \{(i,j) \,;\, \widehat{D}_{i,j} > \xi\}$} with \smash{$X_i$} from \smash{$\mathcal A$} and \smash{$Y_j$} from \smash{$\mathcal B$}. \smash{$Y_j$} equals \smash{$Y_i$} for truly linked pairs but belongs to another entity for falsely linked records. Among the linked data, a proportion \smash{$1-\text{FDP}$} (that may in reality depend on \smash{$X$}) carries the correct \smash{$Y$} and a proportion \smash{$\text{FDP}$} carries a spurious one. Below we denote by \smash{$\widehat{\boldsymbol{\Delta}}$} the estimator of \smash{$\boldsymbol{\Delta}$}, with \smash{$\widehat{\Delta}_{i,j} = 1$} for pairs \smash{$(i,j)$} in \smash{$\mathcal{D}$}. Truly linked pairs of records (true positives, TP) are thus denoted \smash{$\{ \widehat{\boldsymbol{\Delta}} = 1, \boldsymbol{\Delta} = 1 \}$} and falsely linked pairs of records (false positives, FP) \smash{$\{ \widehat{\boldsymbol{\Delta}} = 1, \boldsymbol{\Delta} = 0 \}$}. Heuristically,
\begin{equation}
\begin{aligned}
\mathbb E[ Y \mid X, \widehat{\boldsymbol{\Delta}} = 1 ] &= (1-\text{FDP}) \cdot \mathbb E[ Y \mid X, \widehat{\boldsymbol{\Delta}} = 1, \boldsymbol{\Delta} = 1 ] \\
&\quad + \text{FDP} \cdot \mathbb E[ Y \mid X, \widehat{\boldsymbol{\Delta}} = 1, \boldsymbol{\Delta} = 0] \\
&= (1-\text{FDP}) \cdot \mathbb E_{TP}[ Y \mid X ] + \text{FDP} \cdot \mathbb E_{FP}[ Y \mid X ],
\end{aligned}
\end{equation}
where \smash{$\mathbb E_{TP}$} and \smash{$\mathbb E_{FP}$} denote expectations among truly and falsely linked pairs.

Let us recast \smash{$\widehat{\beta}$} accordingly. If \smash{$X$} has the same mean and variance among all data in \smash{$\mathcal{A}$}, then \smash{$\widehat{\beta}$} is a mixture of \smash{$\widehat{\beta}_{TP}$} and \smash{$\widehat{\beta}_{FP}$}, defined as in \cref{eq: beta-hat-0} for the subsets of TP and FP,
\begin{equation}\label{eq: beta-hat}
\widehat{\beta} = (1-\text{FDP}) \cdot \widehat{\beta}_{TP} + \text{FDP} \cdot \widehat{\beta}_{FP}.
\end{equation}
Linkage errors are often assumed to occur at random, in which case \smash{$\mathbb E_{FP}[ Y \mid X ] = \mathbb E_{FP}[ Y ]$} and \smash{$\widehat{\beta}_{FP} = 0$}. However, the PIVs, often demographic, are likely to be associated with variables of interest in the analysis, so it is more realistic to consider an association between \smash{$X$} and \smash{$Y$} among falsely linked pairs.

The estimation error of the analysis can be decomposed into two parts:
\begin{equation}
\widehat{\beta} - \beta = \underbrace{(\widehat{\beta}_{TP} - \beta)}_{\text{sampling bias}} \;\; + \;\; \underbrace{(\widehat{\beta} - \widehat{\beta}_{TP})}_{\text{contamination bias}}.
\end{equation}
Missed links may cause sample selection and falsely linked data may cause measurement error in the linked variables \citep{Wang_Slawski_linked_2022, reflexRL_2020, CRLrobachetal26}.

\subsubsection{Linkage decision rule trade-off}

Raising the threshold \smash{$\xi$} on the linkage scores reduces the number of falsely linked pairs, which lowers the FDP and hence the contamination bias. Falsely linked data would distort the signal in proportion to the FDP. Indeed, from \cref{eq: beta-hat} we derive:
\begin{equation}\label{eq: beta-contamination}
\underbrace{\widehat{\beta} - \widehat{\beta}_{TP}}_{\text{contamination bias}} = \text{FDP} \, (\widehat{\beta}_{FP} - \widehat{\beta}_{TP}).
\end{equation}

However, raising the threshold \smash{$\xi$} also selects fewer and more atypical records, which may increase the sampling bias due to the implicit record linkage selection. Indeed, the chance of being linked may be related to \smash{$X$}; in the context of linear regression, this yields:
\begin{equation}\label{eq: beta-sampling}
\underbrace{\widehat{\beta}_{TP} - \beta}_{\text{sampling bias}} = \frac{\operatorname{Cov}_{TP}(X,\beta X + \varepsilon)}{\operatorname{Var}_{TP}(X)} - \beta = \frac{\operatorname{Cov}_{TP}(X,\varepsilon)}{\operatorname{Var}_{TP}(X)}.
\end{equation}

These aspects motivate the complementary tools we develop in \textit{FlexRL} to quantify these two biases. Estimating the FDP (see \cref{sec: fdp}) and evaluating the divergence of the linked sample from the data sources (see \cref{sec: representativeness}) over a range of values of \smash{$\xi$} may help users appraise this trade-off.

\subsection{False discovery proportion}\label{sec: fdp}

For the decision rule \smash{$\xi \in [0.5, 1)$}, the FDP is \smash{$\text{FDP}(\xi) = \text{FP}(\xi) / \{\text{TP}(\xi) + \text{FP}(\xi)\}$} where \smash{$\text{TP}(\xi)$} and \smash{$\text{FP}(\xi)$} are the numbers of truly and falsely linked pairs. In real data applications this quantity is unknown. Linkage errors can be estimated from the scores of the linkage model \citep{belin_rubin_1995, measurementError_2019, fastLink_paper_2019}, from synthetic data \citep{linkage_bias_datasynthesis24, FDProbachetal25} or from a separate model under conditions on the distribution of the number of records per entity \citep{sensitivity_2002, sensitivity_2020, DG_esterror22}. \textit{FlexRL} implements the first two estimation procedures. We focus on the FDP because it drives the contamination bias of \cref{eq: beta-contamination}; the sampling bias is addressed by the diagnostics of \cref{sec: representativeness}. The estimators are provided as functions of \smash{$\xi$}, together with the linked sample size available for inference, so that the user can choose a threshold controlling the estimated FDP (and hopefully, the true underlying FDP).

\subsubsection{Score based estimator}

If the linkage model is well calibrated, \smash{$\widehat{D}_{i,j}$} should represent the probability that the pair \smash{$(i,j)$} is a true link, and the expected number of truly linked pairs among linked data is \smash{$\sum_{(i,j)} \widehat{D}_{i,j} \, \mathbf{1} \{\widehat{D}_{i,j} > \xi\}$}. Therefore
\begin{equation}\label{eq: fdp-model}
\widehat{\text{FDP}}(\xi) = 1 - \frac{\sum_{(i,j)} \widehat{D}_{i,j} \, \mathbf{1} \{\widehat{D}_{i,j} > \xi\}}{\sum_{(i,j)} \mathbf{1} \{\widehat{D}_{i,j} > \xi\}}.
\end{equation}
This estimator comes at no extra cost for any method that returns linkage scores, but it is only as good as the linkage model.

Notably, the calibration of the scores and the record linkage performance are affected by missing values and mistakes, the size of the overlap between the sources, the number of entities in the population, the discriminative power of the PIVs, their changes over time, and their marginal and joint distributions. The literature discusses the strengths and limitations of different record linkage paradigms in detail, including comparisons of \textit{FlexRL} and \textit{BRL} across a range of configurations \citep{RLrobachetal25, FDProbachetal25}.

\subsubsection{Synthetic data estimator}

An estimator that does not rely on linkage scores can be computed with synthetic data \citep{FDProbachetal25}. Source \smash{$\mathcal B$} is augmented with \smash{$N_{\text{synth}}$} synthetic records, drawn from a generative model fitted on the PIVs of \smash{$\mathcal B$}---for example \textit{arf} \citep{arf_2023}, \textit{synthpop} \citep{synthpop_2016} or \textit{mice} \citep{mice_2011}. Synthetic records by definition do not form a link in \smash{$\mathcal A$}, so synthetic links can represent false positives when links and non-links have similar distributions \citep{FDProbachetal25}. Then \smash{$\text{FP}_{\text{synth}}(\xi)$} is rescaled by \smash{$\nB / N_{\text{synth}}$}, which follows from equating \smash{$\text{FP}(\xi) / (\nA \cdot N_{\mathcal B})$} with \smash{$\text{FP}_{\text{synth}}(\xi) / (\nA \cdot N_{\text{synth}})$}. As for the score based estimator, \smash{$\text{TP}(\xi) + \text{FP}(\xi)$} corresponds to the number of (real) linked pairs \smash{$N_{\text{real}}(\xi) = \sum_{(i,j)} \mathbf{1} \{\widehat{D}_{i,j} > \xi\}$}:
\begin{equation}\label{eq: fdp-synth}
\widehat{\text{FDP}}_{\text{synth}}(\xi) = \frac{\text{FP}_{\text{synth}}(\xi) \cdot \nB / N_{\text{synth}}}{N_{\text{real}}(\xi)}.
\end{equation}
The procedure is repeated over several synthetic draws and the results are averaged. The estimator does not depend on the linkage model, but it depends on the quality of the data synthesis.

When the two estimators disagree, a practitioner should rely on the one whose assumptions are more plausible. The score based estimator relies on the correct specification of the linkage model, including its specific assumptions. It reflects the performance the model would achieve under correct specification. The synthetic data estimator relies on the fidelity of the synthesiser and on the assumption that links and non-links have similar distributions. Earlier work assessed the robustness of both estimators to the degree of overlap between the data sources, the discriminative power of the PIVs and differences between the PIV distributions of true links and non-links \citep{FDProbachetal25}, but not to deviations from PIV independence. In the SHIW in Emilia-Romagna application, the PIV independence is violated (occupation and sex are associated; see the warning in \cref{sec: prepare}). For models assuming independent PIVs, such as \textit{FlexRL} and Fellegi-Sunter mixture models (see \cref{eq: FS}), we expect the score based estimator to be biased, as its validity rests on the violated assumption. The synthetic data estimator though, should remain valid provided the synthesiser captures the dependencies among PIVs. This is in line with what we observe in the output of \cref{fig: print-output}.

\subsection{Representativeness of the linked sample}\label{sec: representativeness}

As explained in \cref{sec: linkage error}, the linked sample may differ from the data sources in ways the analyst cannot see \citep{reflexRL_2020, Wang_Slawski_linked_2022, CRLrobachetal26}. In particular, we appraise the sampling bias of \cref{eq: beta-sampling} by comparing the distributions of variables of interest between the linked data (`selected' through record linkage) and the original data sources. For any decision rule \smash{$\xi \in [0.5, 1)$}, \textit{FlexRL} compares each source with its linked subset using the following measures, computed on sets of variables \smash{$\mathcal V_{\mathcal A}$} and \smash{$\mathcal V_{\mathcal B}$} available in sources \smash{$\mathcal A$} and \smash{$\mathcal B$} respectively:

\begin{itemize}
\item \textit{Agreement rates}: the proportion of linked pairs whose records agree on \smash{$v$}, for each variable \smash{$v \in \mathcal V_{\mathcal A} \cap \mathcal V_{\mathcal B}$} (such as the PIVs).
\item \textit{Standardised Mean Differences} (SMD): \smash{$(\text{mean}(v_{\text{linked}}) - \text{mean}(v_{\text{source}})) / \text{sd}(v_{\text{source}})$}, for a continuous variable \smash{$v$} in a given set; the same quantity is computed for the indicator of each level when \smash{$v$} is categorical.
\item \textit{Overlap of supports} (support IoU): the Intersection over Union (IoU) of the sets of values taken in the linked sample and in the source, for a variable \smash{$v$} in a given set.
\item \textit{Maximum Mean Discrepancy} (MMD): a multivariate measure of the difference between the linked sample and the source, computed jointly over a given set of variables, with a Gaussian kernel and the median heuristic for the bandwidth \citep{gretton2012}.
\end{itemize}

These quantities are computed over a grid of values of \smash{$\xi$}. They are built on the declared pairs and, when available, their linkage scores; they therefore apply to any linkage method.

In \cref{tbl: demonstration-2} we proceed with the demonstration of \cref{sec: demonstration} showcasing the functionalities of \textit{FlexRL} on a simulation study. The synthetic data FDP estimates (`est. FDP* synth') are not reliable for \textit{fedmatch} and \textit{FlexRL} (dynamics), possibly because of the specific linkage procedures these packages implement. For \textit{FlexRL} (dynamics), the cause is likely the misspecified PIV dynamics model, or the added difficulty of estimating the FDP under modelling of PIV dynamics, since the estimate for \textit{FlexRL} (no dynamics) is reliable.

\begin{table}[htbp]
\centering
\begingroup
\scriptsize
\setlength{\tabcolsep}{2pt}
\begin{tabular}{@{}l ccccccccc@{}}
\toprule
 & Naive & BRL & multilink & fastLink & reclin2 & diyar & fedmatch & \makecell{FlexRL\\(dynamics)} & \makecell{FlexRL\\(no dynamics)} \\
\midrule
TP & 118.40 (5.13) & 80.40 (5.04) & 80.60 (5.95) & 101.70 (6.48) & 114.10 (6.38) & 34.90 (20.24) & 122.10 (6.33) & 114.00 (7.92) & 87.50 (8.30) \\
FP & 83.10 (11.02) & 12.30 (3.23) & 14.20 (4.13) & 43.10 (5.59) & 68.30 (8.82) & 3.30 (2.91) & 125.80 (6.32) & 25.90 (5.15) & 19.00 (4.67) \\
FN & 81.60 (5.13) & 119.60 (5.04) & 119.40 (5.95) & 98.30 (6.48) & 85.90 (6.38) & 165.10 (20.24) & 77.90 (6.33) & 86.00 (7.92) & 112.50 (8.30) \\
sensitivity & 0.59 (0.03) & 0.40 (0.03) & 0.40 (0.03) & 0.51 (0.03) & 0.57 (0.03) & 0.17 (0.10) & 0.61 (0.03) & 0.57 (0.04) & 0.44 (0.04) \\
FDP & 0.41 (0.03) & 0.13 (0.03) & 0.15 (0.04) & 0.30 (0.03) & 0.37 (0.03) & 0.08 (0.05) & 0.51 (0.03) & 0.18 (0.03) & 0.18 (0.04) \\
est. FDP score & NA & 0.19 (0.02) & NA & 0.37 (0.02) & 0.44 (0.03) & NA & 0.10 (0.01) & 0.20 (0.01) & 0.18 (0.02) \\
est. FDP* score & NA & 0.21 (0.02) & NA & 0.39 (0.02) & 0.43 (0.07) & NA & 0.10 (0.01) & 0.21 (0.01) & 0.19 (0.02) \\
est. FDP* synth & NA & 0.18 (0.10) & 0.20 (0.05) & 0.33 (0.07) & 0.45 (0.14) & 0.14 (0.07) & 0.08 (0.03) & 0.37 (0.11) & 0.27 (0.08) \\
max SMD & NA & 0.22 (0.05) & 0.22 (0.06) & 0.17 (0.03) & 0.24 (0.04) & 0.68 (0.31) & 0.15 (0.03) & 0.17 (0.04) & 0.22 (0.04) \\
min support IoU & NA & 0.80 (0.01) & 0.80 (0.01) & 0.80 (0.01) & 0.80 (0.01) & 0.76 (0.10) & 0.82 (0.03) & 0.82 (0.03) & 0.81 (0.02) \\
max MMD & NA & 0.01 (0.00) & 0.01 (0.00) & 0.01 (0.00) & 0.01 (0.00) & 0.08 (0.05) & 0.01 (0.00) & 0.01 (0.00) & 0.01 (0.00) \\
time (secs) & 0.27 (0.04) & 0.97 (0.05) & 18.57 (1.48) & 1.29 (0.24) & 0.29 (0.01) & 1.17 (0.06) & 1.44 (0.01) & 4.59 (0.02) & 3.72 (0.01) \\
memory (Mb) & 1.04 (0.05) & 78.60 (10.08) & 105.68 (10.18) & 24.85 (4.42) & 41.52 (0.18) & 104.31 (10.84) & 42.14 (2.59) & 102.11 (9.03) & 70.18 (6.42) \\
\bottomrule
\end{tabular}
\endgroup
\caption{Linkage performance of competing record linkage methods on simulated data for the default decision rule \smash{$\xi=0.5$}, with linked data quality assessed through \textit{FlexRL}. The * symbol denotes the record linkage augmented task, see \cref{sec: fdp}. We report average performance over 10 replications, with standard deviations in parentheses.}
\label{tbl: demonstration-2}
\end{table}

This short simulation study demonstrates the potential of the \textit{FlexRL} package. Since \textit{FlexRL} provides linked data quality diagnostics for any record linkage method, we report three estimates of the FDP along with measures of the divergence of the linked sample from the data sources. \cref{sec: fdp} presented the FDP estimators implemented in the package and \cref{sec: representativeness} discussed the measures of divergence. The FDP estimators are averaged over 5 procedures in each of the 10 replications.

\subsection{Diagnostics}\label{sec: diagnostics}

\texttt{RL\_diagnostics()} gathers the tools of \cref{sec: fdp} and \cref{sec: representativeness}, summarised in \cref{tbl: diagnostics}, and returns an S3 object of class \texttt{RL\_diagnostics}, with \texttt{print} and \texttt{plot} methods. All measures are computed over a grid of thresholds \smash{$\xi \in [0.5,1)$} (when the linkage method allows it by returning scores) or for a fixed threshold when specified. Each diagnostic can also be computed on its own. We present all commands hereafter but show only some outputs. Further details on these diagnostic tools are given in the package vignette `Linkage and Diagnostics with FlexRL' available through \texttt{vignette(`FlexRL-RL-vignette', `FlexRL')} or the \href{https://robachowyk.github.io/FlexRL/index.html}{online documentation}.

\begin{table}[htbp]
\centering\small
\begin{tabular}{@{}p{24mm} p{63mm} p{40mm}@{}}
\toprule
Diagnostic & Computation & Display \\
\midrule
Convergence & \texttt{plot\_StEM\_convergence()} on the chains returned by \texttt{StEM()} & \texttt{plot(x, `convergence')} \\
Linkage scores & \texttt{plot\_linkage\_scores()} & \texttt{plot(x, `scores')} \\
FDP estimation & \texttt{compute\_RL\_FDP\_score()} (\cref{eq: fdp-model}), \texttt{compute\_augmRL\_FDP\_synth()} (\cref{eq: fdp-synth}) & \texttt{plot(x, `FDP')} \\
Discrepancy measures & \texttt{smd()}, \texttt{support\_iou()}, \texttt{mmd()}, \texttt{RL\_agreement()} & \texttt{plot(x, `discrepancy')} \\
Data comparison & \texttt{plot\_distributions()} on each source and its linked subset & \texttt{plot(x, `distributions')} \\
\bottomrule
\end{tabular}
\caption{Diagnostics performed by \texttt{RL\_diagnostics()}.}
\label{tbl: diagnostics}
\end{table}

The function takes several arguments as input. The record linkage output \texttt{fit}, the data sources \texttt{encodedA} and \texttt{encodedB}, the set of variables of interest to compare \texttt{compare\_vars}, their type \texttt{vars\_type\_cont} (\texttt{TRUE} if continuous) and the \texttt{PIVs} are passed to \texttt{RL\_diagnostics()}, which computes FDP estimates and discrepancy measures that characterise the linked sample across a grid of thresholds \smash{$\xi$}. The \texttt{true\_pairs} enables reporting the true FDP and sensitivity when known.
    
\begin{lstlisting}[language=R]
compare_vars <- c("BirthYear", "Sex", "MaritalStatus",
                  "BirthRegion", "Occupation")

vars_type_cont <- list(BirthYear = TRUE, Sex = FALSE, MaritalStatus = FALSE,
                       BirthRegion = FALSE, Occupation = FALSE)

diag <- RL_diagnostics(fit, prep_data$encodedA, prep_data$encodedB,
                       compare_vars, vars_type_cont, PIVs,
                       true_pairs = prep_data$true_pairs, FDP_estimation = TRUE,
                       RL_method = "FlexRL", data = prep_data)
\end{lstlisting}

The FDP estimation is optional (\texttt{FDP\_estimation = TRUE}) since it can be time consuming. For a given \texttt{RL\_method}, it is repeated \texttt{n\_repeats} times and the estimates are averaged. \texttt{compute\_RL\_FDP\_score()} returns the score based estimate on the original task. \texttt{compute\_augmRL\_FDP\_synth()} adds \texttt{n\_synth} records generated by \texttt{synth\_method} to \smash{$\mathcal B$}, links the augmented sources as an auxiliary task and computes both the score based and the synthetic data estimates on it. Similar score based estimates on the original and augmented tasks support the use of the synthetic data estimator. \cref{tbl: demonstration-2} reported these three estimates as `est. FDP score', `est. FDP* score', `est. FDP* synth'.

The \texttt{plot(x, type)} method for an object of class \texttt{RL\_diagnostics} draws the plots listed in \cref{tbl: diagnostics}, with \texttt{type} being one of \texttt{`convergence'}, \texttt{`scores'}, \texttt{`FDP'}, \texttt{`discrepancy'} and \texttt{`distributions'}. With the `convergence' plots, we could verify that the StEM chains have stabilised.

\begin{lstlisting}[language=R]
plot(diag, `convergence')
\end{lstlisting}

With the `scores' plot, we can analyse the distribution of the linkage scores of the \texttt{fit} as shown in \cref{fig: scores}.

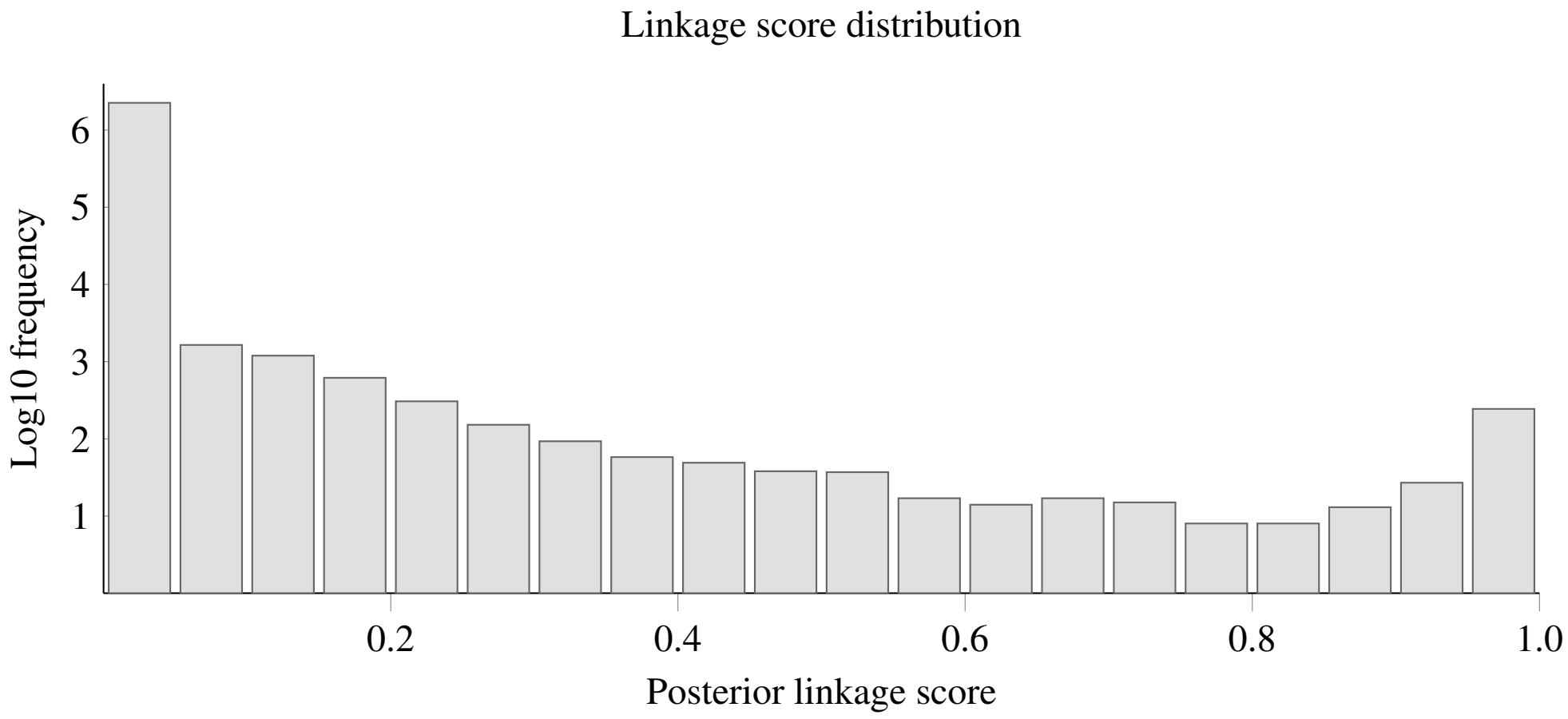
\begin{figure}[htbp]
\centering
\begin{lstlisting}[language=R]
plot(diag, `scores')
\end{lstlisting}
\vspace{1mm}
\begin{tikzpicture}
\begin{axis}[width=0.78\textwidth, height=58mm, ybar, bar width=5.1mm,
    axis lines*=middle,
    xmin=0, xmax=1, ymin=0, ymax=6.6,
    xtick={0,0.2,0.4,0.6,0.8,1.0}, xticklabels={0.0,0.2,0.4,0.6,0.8,1.0},
    ytick={0,1,2,3,4,5,6}, yticklabels={0,1,2,3,4,5,6},
    xlabel={Posterior linkage score}, ylabel={Log10 frequency},
    title={Linkage score distribution}]
\addplot[draw=black!60, fill=black!12] table[x=mid, y=log10] {fig5-scores-hist.dat};
\end{axis}
\end{tikzpicture}
\caption{Distribution of the linkage scores returned by \textit{FlexRL} \texttt{StEM()} on the SHIW in Emilia-Romagna.}
\label{fig: scores}
\end{figure}

Albeit bimodal, the distribution of linkage scores produced by the record linkage algorithm does not indicate where to set the threshold \smash{$\xi$}. The FDP estimation results and the divergence plots may help further. In \cref{fig: tradeoff} we show the `FDP' plot and some of the `discrepancy' plots for the MMD and the SMD. Similar plots for the overlap of supports and the agreement rate are usually displayed by \texttt{plot(diag, `discrepancy')}.

\begin{lstlisting}[language=R]
plot(diag, `FDP')
\end{lstlisting}

\begin{lstlisting}[language=R]
plot(diag, `discrepancy')
\end{lstlisting}

\pgfplotstableread{fig5-fdp-curves.dat}\fdpcurves
\pgfplotstableread{fig5-mmd-curves.dat}\mmdcurves
\pgfplotstableread{fig5-smd-curves.dat}\smdcurves

\begin{figure}[p]
\centering
\begin{tikzpicture}
\begin{axis}[name=fdpL, width=116mm, height=48mm, scale only axis, axis lines=left,
    xmin=0.5, xmax=1.0, ymin=0, ymax=0.26,
    xtick={0.5,0.6,0.7,0.8,0.9,1.0}, xticklabels={0.5,0.6,0.7,0.8,0.9,1.0},
    ytick={0,0.05,0.10,0.15,0.20},
    yticklabels={0.00,0.05,0.10,0.15,0.20},
    xlabel={decision rule threshold}, ylabel={estimated FDP},
    title={FDP versus linkage decision rule},
    legend style={at={(0.985,0.999)}, anchor=north east, draw=none, font=\scriptsize, legend cell align=left, fill=none}]
\addplot[black, thick]        table[x=xi, y=RL_FDP_score]       {\fdpcurves};
\addplot[black, thick, dashed] table[x=xi, y=augmRL_FDP_synth]  {\fdpcurves};
\addplot[black, thick, dotted] table[x=xi, y=augmRL_FDP_score]  {\fdpcurves};
\addplot[black!50]            table[x=xi, y=FDP]                {\fdpcurves};
\addplot[black!30] coordinates {(0.86,0) (0.86,0.26)};
\legend{FDP score (RL), FDP synth (augm. RL), FDP score (augm. RL), true FDP}
\end{axis}
\begin{axis}[at={(fdpL.north east)}, anchor=north east,
    width=116mm, height=48mm, scale only axis,
    axis x line=none, axis y line*=right,
    xmin=0.5, xmax=1.0, ymin=0, ymax=420,
    ytick={0,100,200,300,400}, yticklabels={0,100,200,300,400},
    yticklabel style={color=firebrick},
    y axis line style={firebrick},
    ylabel={number of links},
    ylabel style={color=firebrick},
    legend style={at={(0.699,0.999)}, anchor=north east, draw=none, font=\scriptsize, legend cell align=left, fill=none}]
\addplot[firebrick, thick] table[x=xi, y=RL_n_linked_score]      {\fdpcurves};
\addplot[red, thick]       table[x=xi, y=augmRL_n_linked_synth]  {\fdpcurves};
\addplot[orange, thick]    table[x=xi, y=augmRL_n_linked_score]  {\fdpcurves};
\legend{links (RL), real links (augm. RL), links (augm. RL)}
\end{axis}
\begin{axis}[at={(fdpL.south west)}, anchor=north west, yshift=-30mm,
    width=116mm, height=48mm, scale only axis, axis lines=left,
    xmin=0.5, xmax=1.0, ymin=0.033, ymax=0.080,
    xtick={0.5,0.6,0.7,0.8,0.9,1.0}, xticklabels={0.5,0.6,0.7,0.8,0.9,1.0},
    ytick={0.04,0.05,0.06,0.07},
    yticklabels={0.04,0.05,0.06,0.07},
    scaled y ticks=false,
    yticklabel style={/pgf/number format/fixed, /pgf/number format/precision=2},
    xlabel={decision rule threshold}, ylabel={MMD},
    title={Multivariate discrepancy},
    legend style={at={(0.03,0.985)}, anchor=north west, draw=none, font=\scriptsize, legend cell align=left}]
\addplot[black, thick]        table[x=xi, y=mmdA] {\mmdcurves};
\addplot[black, thick, dashed] table[x=xi, y=mmdB] {\mmdcurves};
\addplot[black!30] coordinates {(0.86,0.033) (0.86,0.080)};
\legend{linked A vs. A, linked B vs. B}
\end{axis}
\begin{axis}[name=smda, at={(fdpL.south west)}, anchor=north west, yshift=-104.8mm,
    width=53mm, height=48mm, scale only axis, axis lines=left,
    xmin=0.5, xmax=1.0, ymin=-1.21, ymax=0.78,
    xtick={0.5,0.6,0.7,0.8,0.9,1.0}, xticklabels={0.5,0.6,0.7,0.8,0.9,1.0},
    ytick={-1.0,-0.5,0,0.5}, yticklabels={$-1.0$,$-0.5$,$0$,$0.5$},
    xlabel={decision rule threshold}, ylabel={SMD},
    title={linked A vs. A},
    ]
\addplot[black]   table[x=xi, y=smdA.MaritalStatus_3]  {\smdcurves};
\addplot[Rtwo]    table[x=xi, y=smdA.MaritalStatus_2]  {\smdcurves};
\addplot[Rthree]  table[x=xi, y=smdA.MaritalStatus_4]  {\smdcurves};
\addplot[Rfour]   table[x=xi, y=smdA.BirthRegion_7]    {\smdcurves};
\addplot[Rfive]   table[x=xi, y=smdA.BirthRegion_13]   {\smdcurves};
\addplot[Rsix]    table[x=xi, y=smdA.BirthRegion_15]   {\smdcurves};
\addplot[Rseven]  table[x=xi, y=smdA.BirthRegion_17]   {\smdcurves};
\addplot[Reight]  table[x=xi, y=smdA.BirthRegion_18]   {\smdcurves};
\addplot[black]   table[x=xi, y=smdA.Occupation_7]     {\smdcurves};
\addplot[Rtwo]    table[x=xi, y=smdA.Occupation_25]    {\smdcurves};
\addplot[Rthree]  table[x=xi, y=smdA.BirthYear]        {\smdcurves};
\addplot[black!30] coordinates {(0.86,-1.21) (0.86,0.78)};
\end{axis}
\begin{axis}[at={(smda.north east)}, anchor=north west, xshift=10mm,
    width=53mm, height=48mm, scale only axis, axis lines=left,
    xmin=0.5, xmax=1.0, ymin=-1.21, ymax=0.78,
    xtick={0.5,0.6,0.7,0.8,0.9,1.0}, xticklabels={0.5,0.6,0.7,0.8,0.9,1.0},
    ytick={-1.0,-0.5,0,0.5}, yticklabels={$-1.0$,$-0.5$,$0$,$0.5$},
    xlabel={decision rule threshold}, ylabel={},
    title={linked B vs. B},
    legend style={at={(0.03,0.985)}, anchor=north west, draw=none, font=\tiny, legend cell align=left}]
\addplot[black]   table[x=xi, y=smdB.MaritalStatus_3]  {\smdcurves};
\addplot[Rtwo]    table[x=xi, y=smdB.MaritalStatus_2]  {\smdcurves};
\addplot[Rthree]  table[x=xi, y=smdB.MaritalStatus_4]  {\smdcurves};
\addplot[Rfour]   table[x=xi, y=smdB.BirthRegion_7]    {\smdcurves};
\addplot[Rfive]   table[x=xi, y=smdB.BirthRegion_13]   {\smdcurves};
\addplot[Rsix]    table[x=xi, y=smdB.BirthRegion_15]   {\smdcurves};
\addplot[Rseven]  table[x=xi, y=smdB.BirthRegion_17]   {\smdcurves};
\addplot[Reight]  table[x=xi, y=smdB.BirthRegion_18]   {\smdcurves};
\addplot[black]   table[x=xi, y=smdB.Occupation_7]     {\smdcurves};
\addplot[Rtwo]    table[x=xi, y=smdB.Occupation_25]    {\smdcurves};
\addplot[Rthree]  table[x=xi, y=smdB.BirthYear]        {\smdcurves};
\addplot[black!30] coordinates {(0.86,-1.21) (0.86,0.78)};
\legend{MaritalStatus\_3, MaritalStatus\_2, MaritalStatus\_4, BirthRegion\_7, BirthRegion\_13, BirthRegion\_15, BirthRegion\_17, BirthRegion\_18, Occupation\_7, Occupation\_25, BirthYear}
\end{axis}
\end{tikzpicture}
\caption{Estimated FDP and discrepancy measures against the linkage decision rule threshold obtained with \textit{FlexRL} \texttt{RL\_diagnostics()} on the SHIW in Emilia-Romagna.}
\label{fig: tradeoff}
\end{figure}
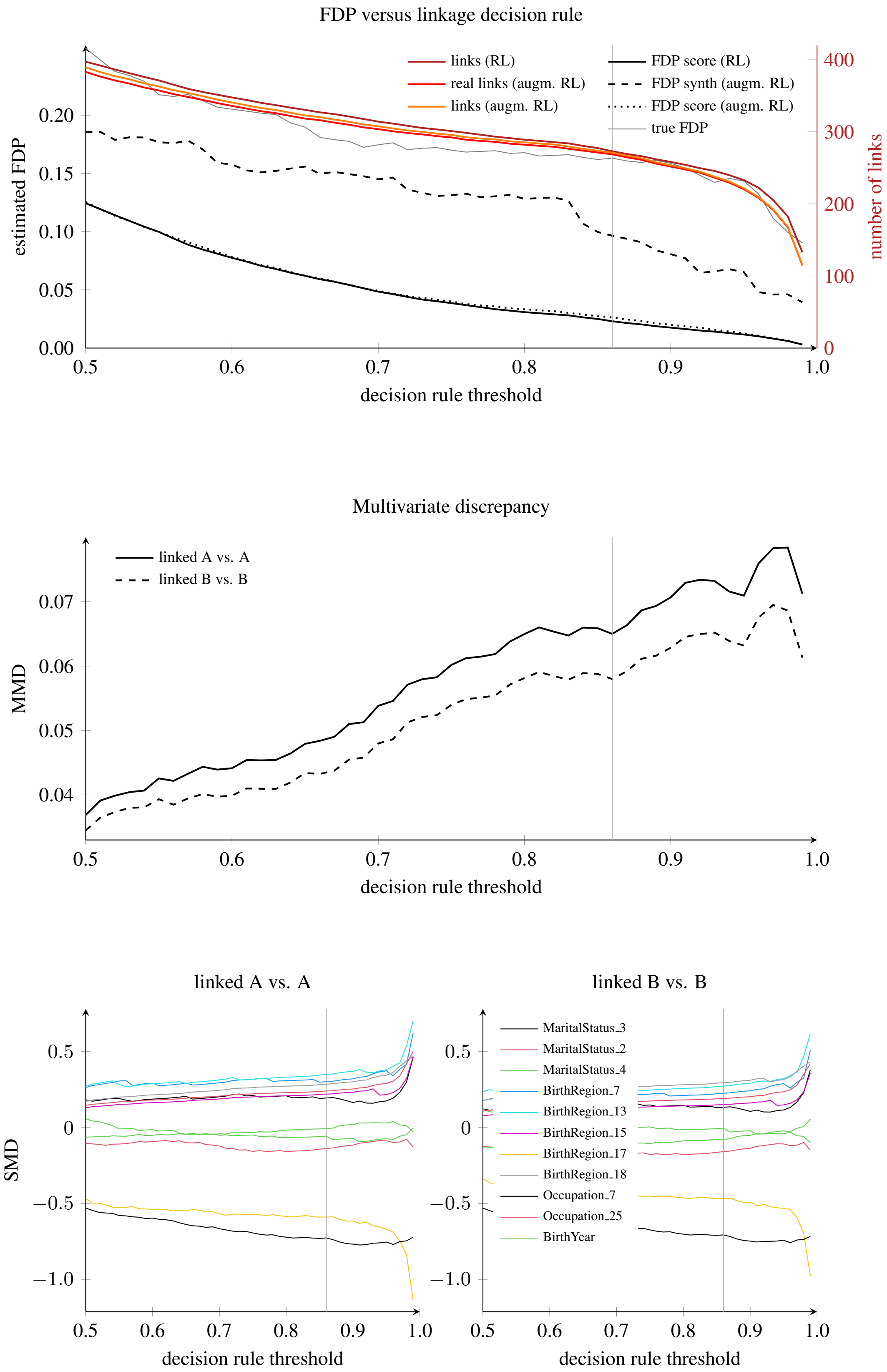

The graphics from \cref{fig: tradeoff} show how each linkage decision rule affects linked data quality. They illustrate the trade-off of \cref{sec: linkage error}: as the threshold \smash{$\xi$} increases, the estimated FDP diminishes while the divergence of the linked sample from the sources grows. Ideally, we choose an appropriate threshold for inference at the `elbow' of the divergence curves, provided that the estimated FDP at that level is reasonable. This decision rule is a heuristic motivated by the potential impact of linkage errors on inference, as discussed in \cref{sec: linkage error}, and its reliability has not been assessed. Note that the divergence measures are indicators of the sampling bias but do not estimate \cref{eq: beta-sampling}. For our example, we choose \smash{$\xi = 0.86$} following this rule, as indicated by the vertical line in \cref{fig: tradeoff}.

With the `distributions' plots, we could compare the distribution of the linked data and of the data sources at the selected threshold---for example \smash{$\xi = 0.86$}.

\begin{lstlisting}[language=R]
plot(diag, `distributions', threshold = 0.86)
\end{lstlisting}

The \texttt{print()} method for an S3 object of class \texttt{RL\_diagnostics} reports the linked sample size, the FDP estimates, the true FDP and sensitivity when available, and the discrepancy measures. We report the output in \cref{fig: print-output}. As discussed in \cref{sec: fdp}, the FDP estimators have different strengths and weaknesses. In the SHIW in Emilia-Romagna, the synthetic data estimator yields an estimate of 0.10 and the score based estimator yields an estimate of 0.02, compared with the true value of 0.16.

\begin{figure}[htbp]
\centering
\begin{lstlisting}[language=R]
print(diag, threshold = 0.86)
\end{lstlisting}
\vspace{1mm}
{\footnotesize
\begin{verbatim}
  Linked pairs  (score > 0.86):  282

  FDP           (true pairs):    0.16
  Sensitivity   (true pairs):    0.26

  FDP score estimate (RL task):            min threshold 0.50: FDP ~ 0.12 (sd 0.00)
                                               threshold 0.86: FDP ~ 0.02 (sd 0.00)
                                           max threshold 0.99: FDP ~ 0.00 (sd 0.00)
  FDP score estimate (augmented RL task):  min threshold 0.50: FDP ~ 0.13 (sd 0.00)
                                               threshold 0.86: FDP ~ 0.03 (sd 0.00)
                                           max threshold 0.99: FDP ~ 0.00 (sd 0.00)
  FDP synth estimate (augmented RL task):  min threshold 0.50: FDP ~ 0.19 (sd 0.06)
                                               threshold 0.86: FDP ~ 0.10 (sd 0.07)
                                           max threshold 0.99: FDP ~ 0.04 (sd 0.07)

  Multivariate MMD  (linked A vs. A):  0.07

  Multivariate MMD  (linked B vs. B):  0.06

  SMD BirthYear                          (linked A vs. A):  -0.01
  SMD Sex values: {1, 2}                 (linked A vs. A):  -0.01, 0.01
  SMD MaritalStatus values: {2, 3, ...}  (linked A vs. A):  -0.14, 0.19, ...
  SMD BirthRegion values: {13, 17, ...}  (linked A vs. A):  0.35, -0.59, ...
  SMD Occupation values: {7, 25, ...}    (linked A vs. A):  -0.73, 0.24, ...

  SMD BirthYear                          (linked B vs. B):  -0.08
  SMD Sex values: {1, 2}                 (linked B vs. B):  -0.01, 0.01
  SMD MaritalStatus values: {2, 3, ...}  (linked B vs. B):  -0.16, 0.13, ...
  SMD BirthRegion values: {17, 18, ...}  (linked B vs. B):  -0.47, 0.30, ...
  SMD Occupation values: {7, 10, ...}    (linked B vs. B):  -0.71, 0.22, ...
\end{verbatim}
}
\caption{Example of output from the \texttt{print()} method on \textit{FlexRL} \texttt{RL\_diagnostics} class, at threshold 0.86, on the SHIW in Emilia-Romagna.}
\label{fig: print-output}
\end{figure}

Throughout the diagnostics, the record linkage is run by wrappers with a common form, \texttt{link\_with\_<pkg>(dataA, dataB, ...)}, returning the indices of linked pairs and their linkage score. Wrappers exist for \textit{FlexRL}, \textit{fastLink}, \textit{BRL}, \textit{reclin2}, \textit{multilink}, \textit{diyar} and \textit{fedmatch}. That way, the diagnostic tools apply to any of these methods in a similar fashion.

On the illustrative example developed along this chapter, we regress the expected age of retirement in 2014 on the expected age of retirement in 2012 (originally named `etapen' in each wave). After preparation of the SHIW in Emilia-Romagna data for record linkage (see \texttt{prepare\_data()} in \cref{sec: prepare}), we work with 1,479 and 1,524 records in the two data sources, which share 898 true links. The solid \textcolor{firebrick}{red} line in \cref{fig: example} is the reference estimate computed on these true links. The tools presented so far help identify a suitable linkage threshold by assessing the trade-off between contamination and sampling bias. In \cref{fig: example}, a low threshold \smash{$\xi=0.55$} yields a linked data estimate (dotted \textcolor{coral1}{orange}) affected by contamination bias, due to noisy falsely linked data (short dashed \textcolor{goldenrod1}{yellow} estimate); see \cref{eq: beta-contamination}. Conversely, a high threshold \smash{$\xi=0.99$} yields a linked data estimate (dotted \textcolor{coral1}{orange}) affected by sampling bias, due to atypical truly linked data (long dashed \textcolor{goldenrod1}{yellow} estimate); see \cref{eq: beta-sampling}. Guided by the diagnostic tools, we choose \smash{$\xi = 0.86$}, which yields a linked data estimate (dotted \textcolor{coral1}{orange}) closer to the reference. At this threshold, we retain 282 linked observations, corresponding to an FDP of 0.16 and a sensitivity of 0.26.

\pgfplotstableread{fig5-example-scatter.dat}\examplepoints

\begin{figure}[p]
\centering
\begin{tikzpicture}
\begin{groupplot}[group style={group size=1 by 3, vertical sep=13mm},
    width=70mm, height=50mm, scale only axis, axis lines=left,
    xmin=60, xmax=80, ymin=60, ymax=80,
    xtick={60,65,70,75,80}, xticklabels={60,65,70,75,80},
    ytick={60,65,70,75,80}, yticklabels={60,65,70,75,80},
    legend style={at={(0.03,0.985)}, anchor=north west, draw=none, font=\scriptsize, legend cell align=left}]
\nextgroupplot[title={linkage decision rule: 0.55}]
\addplot[only marks, mark=*, mark size=1.05pt, black!75, forget plot] table [x=x, y=y] {\examplepoints};
\addplot[firebrick, thick] coordinates {(60, 62.338) (80, 72.269)};
\addplot[coral1, thick, dotted] coordinates {(60, 62.446) (80, 70.988)};
\addplot[goldenrod1, thick, dash pattern=on 5pt off 3pt] coordinates {(60, 62.116) (80, 71.912)};
\addplot[goldenrod1, thick, dashed] coordinates {(60, 64.269) (80, 67.218)};
\legend{true links, {linked data, 363 obs.}, {TP, 284 obs.}, {FP, 79 obs.}}
\node[anchor=south east, font=\scriptsize, align=right] at (axis cs:79.6,60.9) {Sensitivity 0.32\\ FDP 0.22};
\nextgroupplot[title={linkage decision rule: 0.86}, ylabel={expected age of retirement 2014}, yshift={-20pt}]
\addplot[only marks, mark=*, mark size=1.05pt, black!75, forget plot] table [x=x, y=y] {\examplepoints};
\addplot[firebrick, thick] coordinates {(60, 62.338) (80, 72.269)};
\addplot[coral1, thick, dotted] coordinates {(60, 62.494) (80, 71.557)};
\addplot[goldenrod1, thick, dash pattern=on 5pt off 3pt] coordinates {(60, 62.271) (80, 71.906)};
\addplot[goldenrod1, thick, dashed] coordinates {(60, 64.888) (80, 68.199)};
\legend{true links, {linked data, 282 obs.}, {TP, 236 obs.}, {FP, 46 obs.}}
\node[anchor=south east, font=\scriptsize, align=right] at (axis cs:79.6,60.9) {Sensitivity 0.26\\ FDP 0.16};
\nextgroupplot[title={linkage decision rule: 0.99}, xlabel={expected age of retirement 2012}, ylabel={}, yshift={-20pt}]
\addplot[only marks, mark=*, mark size=1.05pt, black!75, forget plot] table [x=x, y=y] {\examplepoints};
\addplot[firebrick, thick] coordinates {(60, 62.338) (80, 72.269)};
\addplot[coral1, thick, dotted] coordinates {(60, 62.550) (80, 70.668)};
\addplot[goldenrod1, thick, dash pattern=on 5pt off 3pt] coordinates {(60, 62.328) (80, 71.307)};
\addplot[goldenrod1, thick, dashed] coordinates {(60, 66.107) (80, 65.188)};
\legend{true links, {linked data, 133 obs.}, {TP, 121 obs.}, {FP, 12 obs.}}
\node[anchor=south east, font=\scriptsize, align=right] at (axis cs:79.6,60.9) {Sensitivity 0.13\\ FDP 0.09};
\end{groupplot}
\end{tikzpicture}
\caption{Regression analysis on the SHIW in Emilia-Romagna linked through \textit{FlexRL}.}
\label{fig: example}
\end{figure}
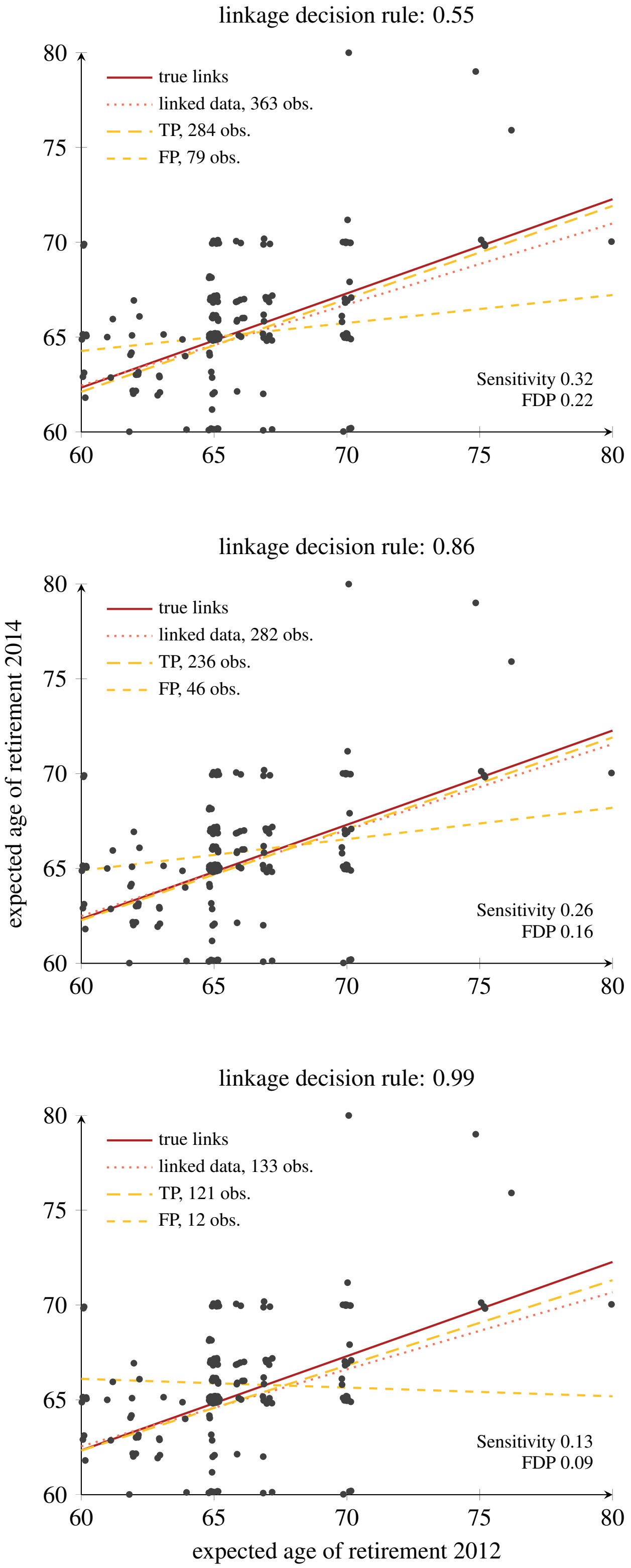

\section{Scalability}\label{sec: scalability}

In \cref{tbl: illustration} of \cref{sec: model}, we compared record linkage methods on the SHIW in Emilia-Romagna. Despite heterogeneous linkage performance, running time and memory usage, all methods ran at that scale. To complete our investigation, we demonstrate the scalability of \textit{FlexRL} and examine that of other record linkage methods on the complete SHIW data sets. After preparation, the 2012 and 2014 waves over the 20 Italian regions cover 19,215 and 20,002 individuals, with 10,568 individuals observed in both waves. We consider two settings, without and with regional blocking. Note that blocking on the region assumes that this variable is stable and free of registration errors.

In \cref{sec: prepare}, we showed the proportion of agreements between the information carried by records among true links of the SHIW in Emilia-Romagna. Below we show the proportion of agreements among true links of the complete SHIW data. In this application, the region is indeed free of registration errors and stable.

{\small
\begin{verbatim}
BirthYear   Sex   MaritalStatus   BirthRegion   Occupation   Region
     0.99  1.00            0.98          1.00         0.82     1.00
\end{verbatim}
}

Record linkage becomes harder on large data sets: the discriminative power of the PIVs is unlikely to grow with the size of the sources, whereas the number of candidate records in \smash{$\mathcal B$} for each record in \smash{$\mathcal A$} does. Without blocking, we may therefore expect higher FDPs and also higher sensitivities, since true links that blocking excludes (due to registration errors or changes between waves) become recoverable. However, in the complete SHIW the region is stable and error free so the expected gain does not apply. FDPs and sensitivities are similar without and with blocking for the comparable methods, except for \textit{FlexRL} which has both a higher FDP and a higher sensitivity without blocking. Note that \texttt{prepare\_data()} eliminates out-of-support records at regional levels, which explains the differences in the partial confusion matrix of the `Naive' method between \cref{tbl: scalability} and \cref{tbl: scalability-block}.

Below, we report the number of unique values of each PIV in the SHIW in Emilia-Romagna and in the complete SHIW. These counts indicate the discriminative power of the PIVs in the regional and national applications. In addition, we appraise the difficulty of the tasks, without and with regional blocking, through the performance of the `Naive' approach in \cref{tbl: scalability} and \cref{tbl: scalability-block}, which reflects how many records share the same combination of PIV values.

{\small
\begin{verbatim}
BirthYear   Sex   MaritalStatus   BirthRegion   Occupation
       94     2               4            18           52
\end{verbatim}
}

{\small
\begin{verbatim}
BirthYear   Sex   MaritalStatus   BirthRegion   Occupation   Region
       99     2               4            20           84       20
\end{verbatim}
}

\cref{tbl: scalability} presents the comparative study without blocking; `Not Available' results indicate memory errors. When supported, methods run with four threads (\texttt{threads} for \textit{FlexRL}, \texttt{n.cores} for \textit{fastLink}, \texttt{nthread} for \textit{fedmatch}). \cref{tbl: scalability-block} shows the results with blocking on the Italian regions. More details on this data application showcasing the scalability of \textit{FlexRL} are given in the package vignette `Scalable Record Linkage with FlexRL' available through \texttt{vignette(`FlexRL-RL-block-vignette', `FlexRL')} or the \href{https://robachowyk.github.io/FlexRL/index.html}{online documentation}.

\begin{table}[htbp]
\centering\small
\setlength{\tabcolsep}{4pt}
\begin{tabular}{@{}l cccccccc@{}}
\toprule
 & Naive & BRL & multilink & fastLink & reclin2 & diyar & fedmatch & FlexRL \\
\midrule
TP & 8,289.00 & NA & NA & 5,061.00 & NA & NA & 5,057.00 & 4,105.00 \\
FP & 49,350.00 & NA & NA & 11,175.00 & NA & NA & 14,154.00 & 5,417.00 \\
FN & 2,279.00 & NA & NA & 5,507.00 & NA & NA & 5,511.00 & 6,463.00 \\
sensitivity & 0.78 & NA & NA & 0.48 & NA & NA & 0.48 & 0.39 \\
FDP & 0.86 & NA & NA & 0.69 & NA & NA & 0.74 & 0.57 \\
time (mins) & 0.00 & NA & NA & 9.00 & NA & NA & 5.00 & 6.00 \\
memory (Mb) & 123.20 & NA & NA & 20,070.10 & NA & NA & 1,877.00 & 4,030.30 \\
\bottomrule
\end{tabular}
\caption{Comparison of record linkage methods on the complete SHIW data application for the default decision rule \smash{$\xi=0.5$}.}
\label{tbl: scalability}
\end{table}

\begin{table}[htbp]
\centering\small
\setlength{\tabcolsep}{2.7pt}
\begin{tabular}{@{}l cccccccc@{}}
\toprule
 & Naive & BRL & multilink & fastLink & reclin2 & diyar & fedmatch & FlexRL \\
\midrule
TP & 8,289.00 & 1,609.00 & 1,771.00 & 4,911.00 & 8,993.00 & 316.00 & 5,284.00 & 2,916.00 \\
FP & 49,346.00 & 449.00 & 1,092.00 & 11,295.00 & 4,092,774.00 & 12.00 & 13,123.00 & 1308.00 \\
FN & 2,059.00 & 8,739.00 & 8,577.00 & 5,437.00 & 1,355.00 & 10,032.00 & 5,064.00 & 7,432.00 \\
sensitivity & 0.80 & 0.16 & 0.17 & 0.48 & 0.87 & 0.03 & 0.51 & 0.28 \\
FDP & 0.86 & 0.22 & 0.38 & 0.70 & 1.00 & 0.04 & 0.71 & 0.31 \\
time (mins) & 0.33 & 3.63 & 206.82 & 5.82 & 0.42 & 2.67 & 2.25 & 5.03 \\
memory (Mb) & 16.00 & 1,150.00 & 3,837.00 & 4,716.00 & 2,223.00 & 1,530.00 & 766.00 & 862.00 \\
\bottomrule
\end{tabular}
\caption{Comparison of record linkage methods on the complete SHIW data application with regional blocking, for the default decision rule \smash{$\xi=0.5$}.}
\label{tbl: scalability-block}
\end{table}

\section{Discussion}\label{sec: discussion}

This chapter presented \textit{FlexRL}: an \texttt{R} package for probabilistic record linkage of two overlapping data sources using a hierarchical model that accounts for registration errors and accommodates temporal changes in Partially Identifying Variables (PIVs). \textit{FlexRL} provides tools for assessing the quality of the resulting linked data. These diagnostics are independent of the linkage method and thus apply broadly. These tools are essential for using record linkage in practice, where potential linkage errors and induced sample selection cannot be observed directly.

To draw attention to the implications of linkage quality for statistical analysis, \textit{FlexRL} provides two complementary tools as functions of the linkage threshold. First, the estimated False Discovery Proportion (FDP), which quantifies contamination among the linked pairs. Second, the divergence of the linked sample from the source populations, which captures potential sampling bias. Examining these quantities jointly makes the trade-off between contamination and sampling bias explicit and provides users with information to select a linkage decision rule in light of its potential consequences for subsequent inference.

Several directions follow naturally from this work. On the computational side, the memory footprint is linear in the number of records and the running time scales with the number of records, candidate pairs, PIVs, their levels and configurations. Memory and running time allow record linkage between large data sets on a standard computer. On the modelling side though, the linkage model currently assumes independence of PIVs given the latent true values, no duplicates within a source, categorical PIVs; extensions would broaden the range of applications. Moreover, the diagnostic estimators and the proposed quality measures, whose validity rests on the assumptions underlying their estimation, would benefit from further theoretical and empirical investigation.

Notwithstanding these considerations, \textit{FlexRL} provides a framework that connects record linkage to linked data quality assessment. It offers a solution to record linkage that is scalable in terms of memory usage and running time, and flexible in accommodating diverse scenarios. Linkage performance depends on the difficulty of the task and the chosen decision rule, controlling the FDP and recovering a fraction of the true links accordingly. It brings record linkage into practice, helping researchers and practitioners not only to link data, but also to quantify and communicate the extent to which linkage decisions may affect downstream inference.

\section{Author contributions statement}
KR developed the record linkage algorithm pioneered by MH and implemented tools for inference conceptualized by MW. All authors drew up the simulation framework. KR conducted the analysis, prepared the first manuscript draft, reviewed and edited by MH and MW. All authors read and approved this manuscript.

\section{Acknowledgments}
The authors thank the anonymous reviewers for their valuable suggestions and for conducting a thorough reproducibility check.

\section{Data and Code Availability}
The \textit{FlexRL} package is available from the Comprehensive R Archive Network (CRAN) or from the package \href{https://github.com/robachowyk/FlexRL/}{GitHub repository} alongside documentation and vignettes describing its functionality. \textit{FlexRL} is also available as a module in JASP 1.0 \citep{JASP2026}. The script is available as supplementary material and on the \href{https://github.com/robachowyk/FlexRL/tree/master/vignettes/article}{GitHub repository} of the package.

\section{Financial disclosure}
This research received no specific grant from any funding agency in the public, commercial, or not-for-profit sectors.

\section{Conflict of interest}
The authors are the developers of \textit{FlexRL}.

\bibliographystyle{abbrvnat}
\bibliography{reference}

\begin{thebibliography}{60}
\providecommand{\natexlab}[1]{#1}
\providecommand{\url}[1]{\texttt{#1}}
\expandafter\ifx\csname urlstyle\endcsname\relax
  \providecommand{\doi}[1]{doi: #1}\else
  \providecommand{\doi}{doi: \begingroup \urlstyle{rm}\Url}\fi

\bibitem[Aleshin-Guendel(2023)]{multilinkmanual}
S.~Aleshin-Guendel.
\newblock \emph{Multifile Record Linkage and Duplicate Detection}, 2023.
\newblock URL \url{https://CRAN.R-project.org/package=multilink}.
\newblock R package version 0.1.1.

\bibitem[Aleshin-Guendel and Sadinle(2023)]{multifilepartitioningrecordlinkage}
S.~Aleshin-Guendel and M.~Sadinle.
\newblock Multifile partitioning for record linkage and duplicate detection.
\newblock \emph{Journal of the American Statistical Association}, 118\penalty0 (543):\penalty0 1786--1795, 2023.
\newblock \doi{10.1080/01621459.2021.2013242}.
\newblock URL \url{https://doi.org/10.1080/01621459.2021.2013242}.
\newblock PMID: 37771512.

\bibitem[Belin and Rubin(1995)]{belin_rubin_1995}
T.~R. Belin and D.~B. Rubin.
\newblock A method for calibrating false-match rates in record linkage.
\newblock \emph{Journal of the American Statistical Association}, 90\penalty0 (430):\penalty0 694--707, 1995.
\newblock ISSN 01621459, 1537274X.
\newblock URL \url{http://www.jstor.org/stable/2291082}.

\bibitem[Binette and Steorts(2022)]{Binette_Steorts_2022}
O.~Binette and R.~C. Steorts.
\newblock (almost) all of entity resolution.
\newblock \emph{Science Advances}, 8\penalty0 (12):\penalty0 eabi8021, 2022.
\newblock \doi{10.1126/sciadv.abi8021}.

\bibitem[Blakely and Salmond(2002)]{sensitivity_2002}
T.~Blakely and C.~Salmond.
\newblock Probabilistic record linkage and a method to calculate the positive predictive value.
\newblock \emph{International Journal of Epidemiology}, 31\penalty0 (6):\penalty0 1246--1252, 12 2002.
\newblock ISSN 0300-5771.
\newblock \doi{10.1093/ije/31.6.1246}.
\newblock URL \url{https://doi.org/10.1093/ije/31.6.1246}.

\bibitem[Bukke et~al.(2026)Bukke, Kamat, Cui, Gutman, and Slawski]{postlink26}
P.~Bukke, G.~Kamat, J.~Cui, R.~Gutman, and M.~Slawski.
\newblock \emph{postlink: Post-Linkage Data Analysis}, 2026.
\newblock URL \url{https://postlink-group.github.io/postlink/}.
\newblock R package version 0.1.2.

\bibitem[Celeux and Diebolt(1986)]{celeux_diebolt_SEM}
G.~Celeux and J.~Diebolt.
\newblock L'algorithme {S}{E}{M} : un algorithme d'apprentissage probabiliste pour la reconnaissance de m\'{e}lange de densit\'{e}s.
\newblock \emph{Revue de {S}tatistiques {A}ppliqu\'{e}es}, 34\penalty0 (2):\penalty0 35--52, 1986.
\newblock URL \url{http://www.numdam.org/item/RSA_1986__34_2_35_0/}.

\bibitem[Cleaton et~al.(2024)Cleaton, Hall, Shipsey, White, and Xhaferaj]{Cleaton_Splink_2024}
M.~Cleaton, J.~Hall, R.~Shipsey, Z.~White, and K.~Xhaferaj.
\newblock A case study of using {S}plink: Census duplicate matching.
\newblock In \emph{Proceedings of Statistics Canada Symposium 2022. Data Disaggregation: Building a More Representative Data Portrait of Society}. Statistics Canada, 2024.

\bibitem[Cohen et~al.(2021)Cohen, Dice, Friedrichs, Gupta, Hayes, Kitschelt, Lee, Marsh, Mislang, Shaton, Sicilian, and Webster]{fedmatch_article}
G.~J. Cohen, J.~Dice, M.~Friedrichs, K.~Gupta, W.~Hayes, I.~Kitschelt, S.~J. Lee, W.~B. Marsh, N.~Mislang, M.~Shaton, M.~Sicilian, and C.~Webster.
\newblock The {U}.{S}. syndicated loan market: Matching data.
\newblock \emph{Journal of Financial Research}, 44\penalty0 (4):\penalty0 695--723, 2021.
\newblock \doi{https://doi.org/10.1111/jfir.12261}.
\newblock URL \url{https://onlinelibrary.wiley.com/doi/abs/10.1111/jfir.12261}.

\bibitem[Cooney et~al.(2023)Cooney, Yu, Daral, Field, Krebs, Kieran, Manley, Govinda~Raju, and Zelt]{cooneyhealthcare_2023}
T.~Cooney, S.~Yu, S.~Daral, M.~Field, B.~Krebs, M.~Kieran, P.~Manley, S.~Govinda~Raju, and S.~Zelt.
\newblock Healthcare resource use for pediatric low-grade glioma care: A cohort of linked electronic health records and claims data.
\newblock \emph{Journal of Clinical Oncology}, 41\penalty0 (16\_suppl):\penalty0 10038--10038, 2023.
\newblock \doi{10.1200/JCO.2023.41.16\_suppl.10038}.
\newblock URL \url{https://ascopubs.org/doi/abs/10.1200/JCO.2023.41.16_suppl.10038}.

\bibitem[Dasylva and Goussanou(2020)]{sensitivity_2020}
A.~Dasylva and A.~Goussanou.
\newblock Estimating linkage errors under regularity conditions.
\newblock In \emph{Proceedings of the Survey Research Methods Section}. American Statistical Association, 2020.

\bibitem[Dasylva and Goussanou(2022)]{DG_esterror22}
A.~Dasylva and A.~Goussanou.
\newblock On the consistent estimation of linkage errors without training data.
\newblock \emph{Japanese Journal of Statistics and Data Science}, 5, 2022.
\newblock \doi{10.1007/s42081-022-00153-3}.

\bibitem[de~Bruin(2019)]{PYRL_de_bruin_j_2019}
J.~de~Bruin.
\newblock Python record linkage toolkit: A toolkit for record linkage and duplicate detection in python, 2019.
\newblock URL \url{https://doi.org/10.5281/zenodo.3559043}.
\newblock Python package.

\bibitem[Doidge and Harron(2020)]{reflexRL_2020}
J.~C. Doidge and K.~L. Harron.
\newblock Reflections on modern methods: linkage error bias.
\newblock \emph{International Journal of Epidemiology}, 04 2020.
\newblock \doi{10.1093/ije/dyz203}.

\bibitem[Dunn(1946)]{bookoflife}
H.~L. Dunn.
\newblock Record linkage.
\newblock \emph{American {J}ournal of {P}ublic {H}ealth and the {N}ations {H}ealth}, 36\penalty0 (12):\penalty0 1412--1416, 1946.

\bibitem[Dusetzina et~al.(2014)Dusetzina, Tyree, Meyer, Meyer, Green, and Carpenter]{review_methods_2014}
S.~B. Dusetzina, S.~Tyree, A.~M. Meyer, A.~Meyer, L.~Green, and W.~R. Carpenter.
\newblock Linking data for health services research: A framework and instructional guide [internet].
\newblock https://www.ncbi.nlm.nih.gov/books/NBK253312/, 2014.
\newblock URL \url{https://www.ncbi.nlm.nih.gov/books/NBK253312/}.
\newblock An Overview of Record Linkage Methods.

\bibitem[Enamorado et~al.(2019)Enamorado, Fifield, and Imai]{fastLink_paper_2019}
T.~Enamorado, B.~Fifield, and K.~Imai.
\newblock Using a probabilistic model to assist merging of large-scale administrative records.
\newblock \emph{American Political Science Review}, 113\penalty0 (2):\penalty0 353–371, 2019.
\newblock \doi{10.1017/S0003055418000783}.

\bibitem[Fellegi and Sunter(1969)]{FS_RL_1969}
I.~P. Fellegi and A.~B. Sunter.
\newblock A theory for record linkage.
\newblock \emph{Journal of the {A}merican {S}tatistical {A}ssociation}, 64\penalty0 (328):\penalty0 1183--1210, 1969.

\bibitem[Fortini et~al.(2001)Fortini, Liseo, Nuccitelli, and Scanu]{fortini_bay_RL_2001}
M.~Fortini, B.~Liseo, A.~Nuccitelli, and M.~Scanu.
\newblock On {B}ayesian record linkage.
\newblock \emph{Research in {O}fficial {S}tatistics}, 4\penalty0 (6):\penalty0 185--198, 2001.

\bibitem[Friedrichs et~al.(2023)Friedrichs, Webster, Marsh, Dice, and Lee]{fedmatchmanual}
M.~Friedrichs, C.~Webster, B.~Marsh, J.~Dice, and S.~Lee.
\newblock \emph{fedmatch: Fast, Flexible, and User-Friendly Record Linkage Methods}, 2023.
\newblock URL \url{https://CRAN.R-project.org/package=fedmatch}.
\newblock R package version 2.0.6.

\bibitem[Gretton et~al.(2012)Gretton, Borgwardt, Rasch, Sch{\"o}lkopf, and Smola]{gretton2012}
A.~Gretton, K.~M. Borgwardt, M.~J. Rasch, B.~Sch{\"o}lkopf, and A.~Smola.
\newblock A kernel two-sample test.
\newblock \emph{Journal of Machine Learning Research}, 13:\penalty0 723--773, 2012.

\bibitem[Harron et~al.(2017)Harron, Doidge, Knight, Gilbert, Goldstein, Cromwell, and van~der Meulen]{eval_linkage_17}
K.~L. Harron, J.~C. Doidge, H.~E. Knight, R.~E. Gilbert, H.~Goldstein, D.~A. Cromwell, and J.~H. van~der Meulen.
\newblock A guide to evaluating linkage quality for the analysis of linked data.
\newblock \emph{International Journal of Epidemiology}, 46\penalty0 (5):\penalty0 1699--1710, 09 2017.
\newblock ISSN 0300-5771.
\newblock \doi{10.1093/ije/dyx177}.
\newblock URL \url{https://doi.org/10.1093/ije/dyx177}.

\bibitem[Harron K~L(2020)]{datalinkagequal2020}
G.~H. Harron K~L, Doidge J~C.
\newblock Assessing data linkage quality in cohort studies.
\newblock \emph{Annals of Human Biology}, 47:\penalty0 218--226, 2020.
\newblock \doi{10.1080/03014460.2020.1742379}.

\bibitem[Jaro(1989)]{jaro_RL_matching_1989}
M.~A. Jaro.
\newblock Advances in record-linkage methodology as applied to matching the 1985 census of tampa, florida.
\newblock \emph{Journal of the {A}merican {S}tatistical {A}ssociation}, 84\penalty0 (406):\penalty0 414--420, 1989.

\bibitem[{JASP Team}(2026)]{JASP2026}
{JASP Team}.
\newblock {JASP (Version 1.0.0)[Computer software]}, 2026.
\newblock URL \url{https://jasp-stats.org/}.

\bibitem[Jelveh et~al.(2022)Jelveh, McNeill, and Lin]{PYRL_namematch}
Z.~Jelveh, M.~McNeill, and T.-Y. Lin.
\newblock Namematch, 2022.
\newblock URL \url{https://github.com/urban-labs/namematch}.
\newblock Python package.

\bibitem[Kramer(2026)]{Kramer_Destatis_2026}
R.~Kramer.
\newblock Towards scalable probabilistic linkage of administrative population data in {G}ermany: Insights from a method test with {S}plink.
\newblock \emph{International Journal of Population Data Science}, 11\penalty0 (4), 2026.
\newblock \doi{10.23889/ijpds.v11i4.3814}.

\bibitem[Lahiri and Larsen(2005)]{Lahiri_Larsen_linked_2005}
P.~Lahiri and M.~D. Larsen.
\newblock Regression analysis with linked data.
\newblock \emph{Journal of the American Statistical Association}, 100\penalty0 (469):\penalty0 222--230, 2005.
\newblock ISSN 01621459.
\newblock URL \url{http://www.jstor.org/stable/27590532}.

\bibitem[Laidler et~al.(2025)Laidler, Imaz~Blanco, and Balasubramanian]{Laidler_NHSE_2025}
J.~Laidler, A.~Imaz~Blanco, and D.~Balasubramanian.
\newblock Probabilistic linkage pipeline improving linkage quality and explainability in healthcare.
\newblock \emph{International Journal of Population Data Science}, 10\penalty0 (4), 2025.
\newblock \doi{10.23889/ijpds.v10i4.3271}.

\bibitem[Lam et~al.(2024)Lam, Boyd, Linacre, Blackburn, and Harron]{linkage_bias_datasynthesis24}
J.~Lam, A.~Boyd, R.~Linacre, R.~Blackburn, and K.~Harron.
\newblock Generating synthetic identifiers to support development and evaluation of data linkage methods.
\newblock \emph{International Journal of Population Data Science}, 9\penalty0 (1), July 2024.
\newblock \doi{10.23889/ijpds.v9i1.2389}.
\newblock URL \url{https://ijpds.org/index.php/ijpds/article/view/2389}.

\bibitem[Larsen(1999)]{Larsen_mixture_1999}
M.~D. Larsen.
\newblock Multiple imputation analysis of records linked using mixture models.
\newblock \emph{Proceedings of the {S}urvey {M}ethods {S}ection, {S}tatistical {S}ociety of {C}anada {A}nnual {M}eeting}, 63:\penalty0 65--71, 1999.

\bibitem[Linacre et~al.(2022)Linacre, Lindsay, Manassis, Slade, Hepworth, Kennedy, and Bond]{splink_2022}
R.~Linacre, S.~Lindsay, T.~Manassis, Z.~Slade, T.~Hepworth, R.~Kennedy, and A.~Bond.
\newblock Splink: Free software for probabilistic record linkage at scale.
\newblock \emph{International Journal of Population Data Science}, 7\penalty0 (3), Aug. 2022.
\newblock \doi{10.23889/ijpds.v7i3.1794}.
\newblock URL \url{https://ijpds.org/article/view/1794}.

\bibitem[Marchant et~al.(2021)Marchant, Kaplan, Elazar, Rubinstein, and Steorts]{dblink_2021}
N.~G. Marchant, A.~Kaplan, D.~N. Elazar, B.~I.~P. Rubinstein, and R.~C. Steorts.
\newblock d-blink: Distributed end-to-end {B}ayesian entity resolution.
\newblock \emph{Journal of {C}omputational and {G}raphical {S}tatistics}, 30\penalty0 (2):\penalty0 406–--421, 2021.

\bibitem[Marchant et~al.(2023)Marchant, Rubinstein, and Steorts]{marchantsteorts2023}
N.~G. Marchant, B.~I.~P. Rubinstein, and R.~C. Steorts.
\newblock Bayesian graphical entity resolution using exchangeable random partition priors.
\newblock \emph{Journal of {S}urvey {S}tatistics and {M}ethodology}, 11\penalty0 (3):\penalty0 569--596, 2023.

\bibitem[Mudgal et~al.(2018)Mudgal, Li, Rekatsinas, Doan, Park, Krishnan, Deep, Arcaute, and Raghavendra]{DeepMatcher_method_2018}
S.~Mudgal, H.~Li, T.~Rekatsinas, A.~Doan, Y.~Park, G.~Krishnan, R.~Deep, E.~Arcaute, and V.~Raghavendra.
\newblock Deep learning for entity matching: A design space exploration.
\newblock In \emph{Proceedings of the 2018 International Conference on Management of Data}, SIGMOD '18, page 19–34, New York, NY, USA, 2018. SIGMOD 18: Proceedings of the 2018 International Conference on Management of Data, Association for Computing Machinery.
\newblock ISBN 9781450347037.
\newblock \doi{10.1145/3183713.3196926}.
\newblock URL \url{https://doi.org/10.1145/3183713.3196926}.
\newblock Code available at \url{https://pypi.org/project/deepmatcher/}.

\bibitem[Newcombe et~al.(1959)Newcombe, Kennedy, Axford, and James]{Newcombe1959}
H.~B. Newcombe, J.~M. Kennedy, S.~J. Axford, and A.~P. James.
\newblock Automatic linkage of vital records.
\newblock \emph{Science}, 130\penalty0 (3381):\penalty0 954--959, 1959.
\newblock \doi{10.1126/science.130.3381.954}.

\bibitem[Nowok et~al.(2016)Nowok, Raab, and Dibben]{synthpop_2016}
B.~Nowok, G.~M. Raab, and C.~Dibben.
\newblock synthpop: Bespoke creation of synthetic data in r.
\newblock \emph{Journal of Statistical Software}, 74\penalty0 (11):\penalty0 1–26, 2016.
\newblock \doi{10.18637/jss.v074.i11}.
\newblock URL \url{https://CRAN.R-project.org/package=synthpop}.

\bibitem[Nsonwu(2023)]{diyarmanual}
O.~Nsonwu.
\newblock \emph{diyar: Record Linkage and Epidemiological Case Definitions in 'R'}, 2023.
\newblock URL \url{https://olisansonwu.github.io/diyar/index.html}.
\newblock R package version 0.5.1.

\bibitem[Pontesilli et~al.(2021)Pontesilli, Hof, Ravelli, van Altena, Soufan, Mol, Kostelijk, Slappendel, Consten, Cantineau, van~der Westerlaken, van Inzen, Dumoulin, Ramos, Baart, Broekmans, Rijnders, Curfs, Mastenbroek, Repping, Roseboom, and Painter]{perined_linkage_2021}
M.~Pontesilli, M.~H. Hof, A.~C.~J. Ravelli, A.~J. van Altena, A.~T. Soufan, B.~W. Mol, E.~H. Kostelijk, E.~Slappendel, D.~Consten, A.~E.~P. Cantineau, L.~A.~J. van~der Westerlaken, W.~van Inzen, J.~C.~M. Dumoulin, L.~Ramos, E.~B. Baart, F.~J.~M. Broekmans, P.~M. Rijnders, M.~H. J.~M. Curfs, S.~Mastenbroek, S.~Repping, T.~J. Roseboom, and R.~C. Painter.
\newblock {Effect of parental and ART treatment characteristics on perinatal outcomes}.
\newblock \emph{Human Reproduction}, 36\penalty0 (6):\penalty0 1640--1665, 04 2021.
\newblock ISSN 0268-1161.
\newblock \doi{10.1093/humrep/deab008}.
\newblock URL \url{https://doi.org/10.1093/humrep/deab008}.

\bibitem[Resnick and Asher(2019)]{measurementError_2019}
D.~Resnick and J.~Asher.
\newblock Measurement of type i and type ii record linkage error.
\newblock In \emph{Using and Linking Administrative and Auxiliary Data}, Government Statistics Section. Joint Statistical Meetings, 09 2019.
\newblock URL \url{https://ww2.amstat.org/meetings/proceedings/2019/data/assets/pdf/1199462.pdf}.

\bibitem[Robach and Hof(2026)]{flexrlmanual}
K.~Robach and M.~H. Hof.
\newblock \emph{FlexRL: Flexible Record Linkage and Inference on Linked Data}, 2026.
\newblock URL \url{https://CRAN.R-project.org/package=flexrl}.
\newblock R package version 2.0.0.

\bibitem[Robach et~al.(2025{\natexlab{a}})Robach, Hof, and van~de Wiel]{FDProbachetal25}
K.~Robach, M.~H. Hof, and M.~A. van~de Wiel.
\newblock False discovery estimation in record linkage.
\newblock \emph{Statistics in Medicine}, 44\penalty0 (23-24):\penalty0 e70292, 2025{\natexlab{a}}.
\newblock \doi{https://doi.org/10.1002/sim.70292}.
\newblock URL \url{https://onlinelibrary.wiley.com/doi/abs/10.1002/sim.70292}.

\bibitem[Robach et~al.(2025{\natexlab{b}})Robach, {van der Pas}, {van de Wiel}, and Hof]{RLrobachetal25}
K.~Robach, S.~L. {van der Pas}, M.~A. {van de Wiel}, and M.~H. Hof.
\newblock A flexible model for record linkage.
\newblock \emph{Journal of the Royal Statistical Society Series C: Applied Statistics}, page qlaf016, Feb. 2025{\natexlab{b}}.
\newblock ISSN 0035-9254.
\newblock \doi{10.1093/jrsssc/qlaf016}.
\newblock URL \url{https://doi.org/10.1093/jrsssc/qlaf016}.

\bibitem[Robach et~al.(2026)Robach, {van de Wiel}, Hof, and {van der Pas}]{CRLrobachetal26}
K.~Robach, M.~A. {van de Wiel}, M.~H. Hof, and S.~L. {van der Pas}.
\newblock Causal record linkage, 2026.
\newblock URL \url{https://arxiv.org/pdf/2611.00760}.

\bibitem[Sadinle(2017)]{sadinle_bay_bipartite_RL_2017}
M.~Sadinle.
\newblock Bayesian estimation of bipartite matchings for record linkage.
\newblock \emph{Journal of the {A}merican {S}tatistical {A}ssociation}, 112\penalty0 (518):\penalty0 600--612, 2017.

\bibitem[Sadinle(2020)]{brlmanual}
M.~Sadinle.
\newblock \emph{Beta Record Linkage}, 2020.
\newblock URL \url{https://CRAN.R-project.org/package=BRL}.
\newblock R package version 0.1.0.

\bibitem[Steorts et~al.(2014{\natexlab{a}})Steorts, Hall, and Fienberg]{smered16software}
R.~Steorts, R.~Hall, and S.~Fienberg.
\newblock {SMERED: A Bayesian Approach to Graphical Record Linkage and De-duplication}.
\newblock In S.~Kaski and J.~Corander, editors, \emph{Proceedings of the Seventeenth International Conference on Artificial Intelligence and Statistics}, volume~33 of \emph{Proceedings of Machine Learning Research}, pages 922--930, Reykjavik, Iceland, 22--25 Apr 2014{\natexlab{a}}. PMLR.
\newblock URL \url{https://proceedings.mlr.press/v33/steorts14.html}.

\bibitem[Steorts(2015)]{steorts2015}
R.~C. Steorts.
\newblock Entity resolution with empirically motivated priors.
\newblock \emph{Bayesian {A}nalysis}, 10\penalty0 (4):\penalty0 849--875, 2015.

\bibitem[Steorts et~al.(2014{\natexlab{b}})Steorts, Ventura, Sadinle, and Fienberg]{blocking_methods_2014}
R.~C. Steorts, S.~L. Ventura, M.~Sadinle, and S.~E. Fienberg.
\newblock \emph{A Comparison of Blocking Methods for Record Linkage}.
\newblock Springer {I}nternational {P}ublishing, 2014{\natexlab{b}}.

\bibitem[Steorts et~al.(2016)Steorts, Hall, and Fienberg]{smered2016}
R.~C. Steorts, R.~Hall, and S.~E. Fienberg.
\newblock A {B}ayesian approach to graphical record linkage and deduplication.
\newblock \emph{Journal of the {A}merican {S}tatistical {A}ssociation}, 111\penalty0 (516):\penalty0 1660--1672, 2016.

\bibitem[Tancredi and Liseo(2011)]{tancredi_bay_RL_2011}
A.~Tancredi and B.~Liseo.
\newblock A hierarchical {B}ayesian approach to record linkage and population size problems.
\newblock \emph{The {A}nnals of {A}pplied {S}tatistics}, 5\penalty0 (2B), 2011.

\bibitem[Tancredi and Liseo(2015)]{Tancredi_Liseo_linked_2015}
A.~Tancredi and B.~Liseo.
\newblock Regression analysis with linked data: problems and possible solutions.
\newblock \emph{Statistica}, 75\penalty0 (1):\penalty0 19–35, Jan. 2015.
\newblock \doi{10.6092/issn.1973-2201/5821}.
\newblock URL \url{https://rivista-statistica.unibo.it/article/view/5821}.

\bibitem[Ted~Enamorado(2023)]{fastLinkmanual}
K.~I. Ted~Enamorado, Ben~Fifield.
\newblock \emph{fastLink: Fast Probabilistic Record Linkage with Missing Data}, 2023.
\newblock URL \url{https://CRAN.R-project.org/package=fastLink}.
\newblock R package version 0.6.1.

\bibitem[Tepping(1968)]{Tepping1968}
B.~J. Tepping.
\newblock A model for optimum linkage of records.
\newblock \emph{Journal of the {A}merican {S}tatistical {A}ssociation}, 63:\penalty0 1321--1332, 1968.

\bibitem[van Buuren and Groothuis-Oudshoorn(2011)]{mice_2011}
S.~van Buuren and K.~Groothuis-Oudshoorn.
\newblock mice: Multivariate imputation by chained equations in r.
\newblock \emph{Journal of Statistical Software}, 45\penalty0 (3):\penalty0 1–67, 2011.
\newblock \doi{10.18637/jss.v045.i03}.
\newblock URL \url{https://CRAN.R-project.org/package=mice}.

\bibitem[{van der Laan}(2025)]{reclin2manual}
J.~{van der Laan}.
\newblock \emph{reclin2: Record Linkage Toolkit}, 2025.
\newblock URL \url{https://github.com/djvanderlaan/reclin2}.
\newblock R package version 0.6.0.

\bibitem[Vo et~al.(2024)Vo, Garès, Zhang, Happe, Oger, Paquelet, and Chauvet]{CoxLinkedData_2024}
T.~H. Vo, V.~Garès, L.-C. Zhang, A.~Happe, E.~Oger, S.~Paquelet, and G.~Chauvet.
\newblock Cox regression with linked data.
\newblock \emph{Statistics in Medicine}, 43\penalty0 (2):\penalty0 296--314, 2024.
\newblock \doi{https://doi.org/10.1002/sim.9960}.
\newblock URL \url{https://onlinelibrary.wiley.com/doi/abs/10.1002/sim.9960}.

\bibitem[Wang et~al.(2022)Wang, Ben-David, Diao, and Slawski]{Wang_Slawski_linked_2022}
Z.~Wang, E.~Ben-David, G.~Diao, and M.~Slawski.
\newblock Regression with linked datasets subject to linkage error.
\newblock \emph{WIREs Computational Statistics}, 14\penalty0 (4):\penalty0 e1570, 2022.
\newblock \doi{https://doi.org/10.1002/wics.1570}.
\newblock URL \url{https://wires.onlinelibrary.wiley.com/doi/abs/10.1002/wics.1570}.

\bibitem[Watson et~al.(2023)Watson, Blesch, Kapar, and Wright]{arf_2023}
D.~S. Watson, K.~Blesch, J.~Kapar, and M.~N. Wright.
\newblock Adversarial random forests for density estimation and generative modeling.
\newblock https://arxiv.org/abs/2205.09435, 2023.
\newblock URL \url{https://CRAN.R-project.org/package=arf}.

\bibitem[Williamson et~al.(2020)Williamson, Walker, Bhaskaran, Bacon, Bates, Morton, Curtis, Mehrkar, Evans, Inglesby, Cockburn, McDonald, MacKenna, Tomlinson, Douglas, Rentsch, Mathur, Wong, Grieve, Harrison, Forbes, Schultze, Croker, Parry, Hester, Harper, Perera, Evans, Smeeth, and Goldacre]{opensafely2020}
E.~J. Williamson, A.~J. Walker, K.~Bhaskaran, S.~Bacon, C.~Bates, C.~E. Morton, H.~J. Curtis, A.~Mehrkar, D.~Evans, P.~Inglesby, J.~Cockburn, H.~I. McDonald, B.~MacKenna, L.~Tomlinson, I.~J. Douglas, C.~T. Rentsch, R.~Mathur, A.~Y.~S. Wong, R.~Grieve, D.~Harrison, H.~Forbes, A.~Schultze, R.~Croker, J.~Parry, F.~Hester, S.~Harper, R.~Perera, S.~J.~W. Evans, L.~Smeeth, and B.~Goldacre.
\newblock Opensafely: factors associated with covid-19-related hospital death in the linked electronic health records of 17 million adult nhs patients.
\newblock \emph{MedRxiv}, pages 430--436, 2020.

\end{thebibliography}

\end{document}